\newcommand{\mii}{Mg\,\textsc{ii}}
\newcommand{\civ}{C\,\textsc{iv}}

\documentclass{ws-ijmpd}
\usepackage{hyperref}
\usepackage{url}
\usepackage[super,compress]{cite}
\begin{document}

\markboth{Park \& Ratra}
{Is excess smoothing of Planck data partially responsible for DE dynamics in $w(z)$CDM?}

%
\catchline{}{}{}{}{}
%

\title{Is excess smoothing of Planck CMB ansiotropy data partially responsible for evidence for dark energy dynamics in other $w(z)$CDM parametrizations?}

\author{Chan-Gyung Park}
\address{Division of Science Education and Institute of Fusion Science, Jeonbuk National University,\\
Jeonju 54896, Republic of Korea\\
park.chan.gyung@gmail.com}

\author{Bharat Ratra}
\address{Department of Physics, Kansas State University, 116 Cardwell Hall, Manhattan, KS 66506, USA\\
ratra@ksu.edu}

\maketitle

\begin{history}
\received{Day Month Year}
\revised{Day Month Year}
\end{history}

\begin{abstract}
We study spatially-flat dynamical dark energy parametrizations, $w(z)$CDM, with redshift-dependent dark energy equation of state parameter $w(z)$ expressed using three different quadratic and other polynomial forms (as functions of $1-a$, where $a$ is the scale factor), without and with a varying cosmic microwave background (CMB) lensing consistency parameter $A_L$. We use Planck CMB anisotropy data (P18 and lensing) and a large, mutually-consistent non-CMB data compilation that includes Pantheon+ type Ia supernova, baryon acoustic oscillation (BAO), Hubble parameter ($H(z)$), and growth factor ($f\sigma_8$) measurements, but not recent DESI BAO data. The six $w(z)$CDM ($+A_L$) parametrizations show higher consistency between the CMB and non-CMB data constraints compared to the XCDM ($+A_L$) and $w_0 w_a$CDM ($+A_L$) cases. Constraints from the most-restrictive P18+lensing+non-CMB data compilation on the six $w(z)$CDM ($+A_L$) parametrizations indicate that dark energy dynamics is favored over a cosmological constant by $\gtrsim 2\sigma$ when $A_L = 1$, but only by $\gtrsim 1\sigma$ when $A_L$ is allowed to vary (and $A_L>1$ at $\sim2\sigma$ significance). 
Non-CMB data dominate the P18+lensing+non-CMB compilation at low $z$ and favor quintessence-like dark energy. At high $z$ P18+lensing data dominate, favoring phantom-like dark energy with significance from $1.5\sigma$ to $2.9 \sigma$ when $A_L = 1$, and from $1.1\sigma$ to $1.8\sigma$ when $A_L$ varies. These results suggest that the observed excess weak lensing smoothing of some of the Planck CMB anisotropy multipoles is partially responsible for the $A_L = 1$ cases $\gtrsim 2\sigma$ evidence for dark energy dynamics over a cosmological constant.
\end{abstract}

\ccode{PACS numbers: 98.80.-k}


\section{Introduction}
\label{sec:Introduction} 

In the standard spatially-flat $\Lambda$CDM cosmological model \cite{Peebles:1984ge}, dark energy in the form of a cosmological constant $\Lambda$ dominates the current cosmological energy budget and is responsible for the observed, low-redshift, accelerated cosmological expansion, with cold dark matter (CDM) being the next largest contributor to the current energy budget. This flat $\Lambda$CDM model agrees with many of the observational constraints, but there are some clouds, for recent reviews see Refs.\ \citen{Perivolaropoulos:2021jda, Moresco:2022phi, Abdalla:2022yfr, Hu:2023jqc}. 

A recent example is the DESI Collaboration result \cite{DESI:2024mwx} determined from observational data analyzed in the spatially-flat $w_0w_a$CDM parametrization in which dynamical dark energy is taken to be a fluid with a parameterized equation of state parameter (the ratio of the fluid pressure to the fluid energy density) $w(a) = w_0 + w_a (1 - a) = w_0 + w_a z/(1+z) = w(z)$, a function of redshift $z$ or cosmological scale factor $a$ \cite{Chevallier:2000qy, Linder:2002et}. The analysis of new DESI baryon acoustic oscillation (BAO) measurements in combination with cosmic microwave background (CMB) data and Type Ia supernova (SNIa) observations, in particular the DESI+CMB+PantheonPlus data compilation, see Ref.\ \refcite{DESI:2024mwx} for a detailed description, indicates that these data favor a region in parameter space that is $\gtrsim 2\sigma$ away from the flat $\Lambda$CDM model point at $w_0 = -1$ and $w_a = 0$ \cite{DESI:2024mwx}, thus indicating $\gtrsim 2\sigma$ support for dynamical dark energy over a cosmological constant. For discussions of the DESI result, see Refs.\ \citen{Tada:2024znt, Yin:2024hba, Wang:2024hks, Luongo:2024fww, Cortes:2024lgw, Colgain:2024xqj, Wang:2024rjd, Berghaus:2024kra, Wang:2024dka, Yang:2024kdo, Chan-GyungPark:2024mlx, Shlivko:2024llw, Huang:2024qno, DESI:2024aqx, Dinda:2024kjf, Andriot:2024jsh, DESI:2024kob, Bhattacharya:2024hep, Ramadan:2024kmn, Mukherjee:2024ryz, Roy:2024kni, Wang:2024hwd, Heckman:2024apk, Gialamas:2024lyw, Notari:2024rti, Liu:2024gfy, Orchard:2024bve, Patel:2024odo, Wang:2024sgo, Li:2024qso, Giare:2024gpk, Dinda:2024ktd, Jiang:2024viw, Jiang:2024xnu, Alfano:2024jqn, Ghosh:2024kyd, Pourojaghi:2024bxa, Reboucas:2024smm, Wolf:2024eph, RoyChoudhury:2024wri, Lu:2024hvv, Chan-GyungPark:2024brx, Li:2024qus, Payeur:2024dnq, Ishak:2024jhs, Akthar:2024tua, Gao:2024ily, Tang:2024lmo, Berbig:2024aee} and references therein.

In the $w_0w_a$CDM parametrization the $\gtrsim 2\sigma$ support for dynamical dark energy over a cosmological constant does not require including DESI BAO measurements or SNIa data in the analysis \cite{Chan-GyungPark:2024mlx}. A large compilation of independent, mutually-consistent, non-CMB measurements \cite{deCruzPerez:2024abc} used jointly with Planck CMB anisotropy data \cite{Planck:2018vyg} provide slightly more restrictive constraints and support for dark energy dynamics \cite{Chan-GyungPark:2024mlx} than found from the DESI+CM+PantheonPlus compilation \cite{DESI:2024mwx}. And there have been earlier suggestions that dynamical dark energy is mildly favored over a cosmological constant, see Refs.\ \citen{Sola:2016hnq, Ooba:2017lng, Ooba:2018dzf, Park:2018fxx, SolaPeracaula:2018wwm, Park:2019emi, Khadka:2020vlh, Cao:2020jgu, Khadka:2020hvb, Cao:2021ldv, Cao:2021irf, Dong:2023jtk, VanRaamsdonk:2023ion, VanRaamsdonk:2024sdp, Thompson:2024nxf} and references therein.

While the $\gtrsim 2\sigma$ support for dark energy dynamics in the $w_0w_a$CDM parametrization does not requiring using DESI BAO data or SNIa data in the analysis \cite{Chan-GyungPark:2024mlx}, part of this support seems to be related to the observed excess weak lensing smoothing of some Planck CMB anisotropy multipoles (relative to what is predicted in the Planck best-fit cosmological model) \cite{Calabreseetal2008, Planck:2018vyg}, as follows, Ref.\ \refcite{Chan-GyungPark:2024brx}. Including the lensing consistency parameter $A_L$ \cite{Calabreseetal2008} in the analysis, as a new free parameter to be determined from the dataset being analyzed, allows for a consistency check on the excess weak lensing smoothing. If the dataset under analysis gives a value of $A_L$ that is consistent with unity then there is no excess smoothing. In the standard flat $\Lambda$CDM model, Planck CMB anisotropy data give $A_L > 1$ at $\sim 2\sigma$, indicating that there is excess smoothing \cite{Planck:2018vyg, deCruzPerez:2022hfr}. Including a varying $A_L$ in an analysis of Planck CMB data and non-CMB data in the $w_0w_a$CDM$+A_L$ parametrization \cite{Chan-GyungPark:2024brx}: makes the Planck CMB data constraints and the non-CMB data constraints more consistent; again results in an $A_L > 1$ at $\sim 2\sigma$; and, reduces the support for dark energy dynamics over a cosmological constant to $\gtrsim 1\sigma$, compared to the $A_L = 1$ case support of $\gtrsim 2\sigma$, thus suggesting that the observed excess smoothing of some Planck CMB anisotropy multipoles contributes to the $\gtrsim 2\sigma$ support for dark energy dynamics in the $A_L = 1$ $w_0w_a$CDM parametrization.      

In this study we extend earlier work \cite{Chan-GyungPark:2024mlx, Chan-GyungPark:2024brx} by exploring three new spatially-flat $w(z)$CDM ($+A_L$) dynamical dark energy parametrizations with $w(z)$ expressed in quadratic and other polynomial forms as a function of $1-a$. Utilizing the largest independent and mutually-consistent compilation of non-CMB data to date \cite{deCruzPerez:2024abc} together with Planck CMB data, we study these parametrizations to address the important question of whether dark energy exhibits dynamics beyond the cosmological constant. Through a combined analysis of the CMB (P18 and lensing) and non-CMB data compilation, we observationally constrain these dynamical dark energy parametrizations (without and with a varying CMB lensing consistency parameter $A_L$) and quantitatively investigate how well they fit the observations and how consistent the CMB and non-CMB data constraints are with each other in a given parametrization. We find from results of the analyses of the six $w(z)$CDM ($+A_L$) parametrizations (including $A_L=1$ and varying $A_L$ cases) that there is better consistency between the CMB and non-CMB data constraints than in the simpler XCDM ($+A_L$) parametrizations with constant dark energy equation of state and in the $w_0 w_a$CDM ($+A_L$) parametrizations. In all six cases, the best-fit dark energy is dynamical and has quintessence-like behavior at low $z$ and phantom-like behavior at high $z$. In the $A_L = 1$ cases dark energy dynamics is favored over a cosmological constant at $\gtrsim 2\sigma$, however in the varying $A_L$ cases this support reduces to $\gtrsim 1\sigma$ (with $A_L > 1$ at $\sim2\sigma)$. This again suggests that the observed excess smoothing of some Planck CMB anisotropy multipoles contributes to the $\gtrsim 2\sigma$ support for dark energy dynamics in the three new $A_L = 1$ $w(z)$CDM parametrizations we study here, consistent with the $w_0w_a$CDM($+A_L$) findings of Ref.\ \refcite{Chan-GyungPark:2024brx}. 

In Sec.\ \ref{sec:Data} we provide brief details of the data sets we use to constrain cosmological parameters in, and test the performance of, the flat $w(z)$CDM parametrizations. In Sec.\ \ref{sec:Methods} we briefly summarize the main features of the three flat $w_0w_2$CDM, $w_0 w_p p$CDM, and $w_0 w_1 w_2$CDM parametrizations we study and the analysis techniques we use. Our results are presented and discussed in Sec.\ \ref{sec:ResultsandDiscussion}, and we conclude in Sec.\ \ref{sec:Conclusion}.

\section{Data}
\label{sec:Data}

In this paper CMB and non-CMB data sets are used to constrain the parameters of dynamical dark energy parameterizations. The data sets we use for this purpose are described in detail in Sec.\ II of Ref.\ \refcite{deCruzPerez:2024abc} and outlined in what follows. We account for all known data covariances. 

For the CMB data, we use the Planck 2018 TT,TE,EE+lowE (P18) CMB temperature and polarization power spectra alone as well as jointly with the Planck lensing potential (lensing) power spectrum \cite{Planck:2018nkj,Planck:2018vyg}. 

The non-CMB data set we use is the non-CMB (new) data compilation of Ref.\ \refcite{deCruzPerez:2024abc}, which is comprised of 

\begin{itemize}

\item 16 BAO data points, spanning $0.122 \le z \le 2.334$, listed in Table I of Ref.\ \refcite{deCruzPerez:2024abc}. We do not use DESI 2024 BAO data \cite{DESI:2024mwx}.

\item A 1590 SNIa data point subset of the Pantheon+ compilation \cite{Brout:2022vxf}, retaining only SNIa with $z > 0.01$ to mitigate peculiar velocity correction effects. These data span $0.01016 \le z \le 2.26137$,

\item 32 Hubble parameter [$H(z)$] measurements, spanning $0.070 \le z \le 1.965$, listed in Table 1 of Ref.\ \refcite{Cao:2023eja} and in Table II of Ref.\ \refcite{deCruzPerez:2024abc}. 

\item An additional nine (non-BAO) growth rate ($f\sigma_8$) data points, 
spanning $0.013 \le z \le 1.36$, listed in Table III of Ref.\ \refcite{deCruzPerez:2024abc}.

\end{itemize}

We use five individual and combined data sets to constrain the flat $w(z)$CDM models with three different dark energy equation of state parametrizations: P18 data, P18+lensing data, non-CMB data, P18+non-CMB data, and P18+lensing+non-CMB data.

\section{Methods}
\label{sec:Methods}

Here we summarize the method we applied in this study. Additional details about the methodology we use are described in Sec.\ III of Ref.\ \refcite{deCruzPerez:2024abc}. 

To determine quantitatively how restrictively these observational data constrain cosmological model parameters, we make use of the \texttt{CAMB}/\texttt{COSMOMC} program (October 2018 version) \cite{Challinor:1998xk,Lewis:1999bs,Lewis:2002ah}. \texttt{CAMB} computes the evolution of spatial inhomogeneities and makes theoretical predictions which depend on the cosmological parameters of the dynamical dark energy parameterizations we study here. \texttt{COSMOMC} uses the Markov chain Monte Carlo (MCMC) method to compare these theoretical model predictions to observational data, to determine cosmological parameter likelihoods. The MCMC chains are assumed to have converged when the Gelman and Rubin $R$ statistic (provided by \texttt{COSMOMC}) satisfies $R-1 < 0.01$. For each model and data set, we use the converged MCMC chains, with the \texttt{GetDist} code \cite{Lewis:2019xzd}, to compute the average values, confidence intervals, and likelihood distributions of model parameters. 

To establish a baseline for comparison, we also study the spatially-flat $\Lambda$CDM model. The six primary cosmological parameters for this model are conventionally chosen to be the current value of the physical baryonic matter density parameter $\Omega_b h^2$, the current value of the physical cold dark matter density parameter $\Omega_c h^2$, the angular size of the sound horizon at recombination $100\theta_{\text{MC}}$, the optical depth to reionization $\tau$, the scalar-type primordial perturbation power spectral index $n_s$, and the power spectrum amplitude $\ln(10^{10}A_s)$, where $h$ is the Hubble constant in units of 100 km s$^{-1}$ Mpc$^{-1}$. We assume flat priors for these parameters, non-zero over: $0.005 \le \Omega_b h^2 \le 0.1$, $0.001 \le \Omega_c h^2 \le 0.99$, $0.5 \le 100\theta_\textrm{MC} \le 10$, $0.01 \le \tau \le 0.8$, $0.8 \le n_s \le 1.2$, and $1.61 \le \ln(10^{10} A_s) \le 3.91$. In the flat $\Lambda$CDM$+A_L$ model (and the $w(z)$CDM$+A_L$ parametrizations), for the lensing consistency parameter $A_L$ we adopt a flat prior non-zero over $0 \le A_L \le 10$.

In the $w_0 w_a$CDM parameterization dynamical dark energy is assumed to be a fluid with an evolving equation of state parameter (fluid pressure to energy density ratio) $w(a) = w_0 + w_a (1-a)$ as a function of scale factor $a$ or equivalently $w(z) = w_0 + w_az/(1+z)$ as a function of redshift $z$ \cite{Chevallier:2000qy, Linder:2002et}. Since this two parameter dynamical dark energy parameterization explores a specific dark energy equation of state parameter region, we also study an extended three parameter $w_0w_1w_2$CDM parametrization that includes the quadratic term, 
\begin{equation}
   w(a)=w_0 + w_1 (1-a) + w_2 (1-a)^2 .
\end{equation}
When $w_2 = 0$ and with $w_1 = w_a$, this reduces to the $w_0 w_a$CDM parameterization. (Higher order terms are also possible \cite{Dai:2018zwv}, but current data do not as effectively constrain dynamical dark energy parametrizations with four or more free parameters.) We first explore the two parameter $w_0w_2$CDM case with $w_1=0$, and then the more general case where $w_0$, $w_1$, and $w_2$ all vary.

We also consider an extension of the quadratic ($w_1=0$) case to an  arbitrary order $p$ form, with the three parameter $w_0w_pp$CDM parameterization,
\begin{equation}
   w(a)=w_0 + w_p (1-a)^{p},
\end{equation}
where the order $p$ is an additional dark energy parameter, in addition to $w_0$ and $w_p$. 

In all cases we consider here and in Refs.\ \citen{Chan-GyungPark:2024mlx, Chan-GyungPark:2024brx}, at low redshift these $w(z)$CDM parametrizations behave like an XCDM parametrization with $w_0 = w(z=0)$, while at high redshift they also behave like an XCDM parameterization but now with $w(z \rightarrow \infty) = w_0 + w_a$ (for $w_0w_a$CDM), $= w_0 + w_2$ (for $w_0w_2$CDM), $= w_0 + w_p$ (for $w_0w_pp$CDM), and $= w_0 + w_1 +w_2$ (for $w_0w_1w_2$CDM). Compared to XCDM, these two- and three-parameter $w(z)$CDM parametrizations have more flexibility in fitting data, allowing $w(z=0)$ to better fit low-redshift non-CMB data while $w(z \rightarrow \infty)$ can now better accommodate high-redshift CMB data. To allow for dynamical dark energy parametrizations whose equation of state parameter crosses $w=-1$, the \texttt{CAMB}/\texttt{COSMOMC} option of the parametrized post-Friedmann dark energy model \cite{Fang:2008sn} is used.

For the dynamical dark energy equation of state parameters in the $w_0 w_a$CDM model parameterization we adopt flat priors non-zero over $-3.0 \le w_0 \le 0.2$ and $-3 < w_a < 2$. For the $w_0 w_p p$CDM parameterization, we apply an additional flat prior for the $p$ parameter non-zero over $0 \le p \le 4$ while the flat prior for $w_p$ is non-zero over $-3 \le w_p \le 2$. In the $w_0 w_1 w_2$ CDM parameterization we adopt an additional flat prior non-zero over $-3 \le w_2 \le 2$ with $w_1$ having the same prior as $w_a$. 

When we estimate parameters using non-CMB data, we fix the values of $\tau$ and $n_s$ to those obtained from P18 data (since these parameters cannot be determined solely from non-CMB data) and constrain the other cosmological parameters. 

Additionally, we also present constraints on three derived parameters: the Hubble constant $H_0$, the current value of the non-relativistic matter density parameter $\Omega_m$, and the amplitude of matter fluctuations $\sigma_8$, that are determined from those on the primary parameters of the cosmological model. We also record the values of the sum of dark energy equation of state parameters to which $w(z)$ approaches at high $z$, in the different dynamical dark energy parametrizations we study: $w_0 + w_a$, $w_1 + w_2 + w_3$, $w_0 + w_2$, and $w_0 + w_p$. These are determined from the primary parameters of the corresponding dynamical dark energy parametrization. 

All the $w(z)$CDM parametrizations we study here, as well as the $\Lambda$CDM model, have flat spatial hypersurfaces and a tilted scalar-type primordial scalar-type energy density perturbation power spectrum
\begin{equation}
    P_\delta (k) = A_s \left( \frac{k}{k_0} \right)^{n_s},
\label{eq:powden-flat}
\end{equation}
where $k$ is the wavenumber and $n_s$ and $A_s$ are the spectral index and the amplitude and the power spectrum at pivot scale $k_0=0.05~\textrm{Mpc}^{-1}$. This power spectrum is sourced by zero-point quantum fluctuations during an early epoch of power-law inflation in a spatially-flat inflation model with a scalar field inflaton potential energy density that is an exponential function of the inflaton \cite{Lucchin:1984yf, Ratra:1989uv, Ratra:1989uz}.

\begin{table}[htpb]
\tbl{Consistency check parameter $\log_{10} \mathcal{I}$ and tension parameters $\sigma$ and $p$ (\%) for P18 vs.\ non-CMB datasets and P18+lensing vs.\ non-CMB datasets in the flat $w_0 w_2$CDM, $w_0 w_2$CDM$+A_L$, $w_0 w_p p$CDM, $w_0 w_p p$CDM$+A_L$, $w_0 w_1 w_2$CDM, and $w_0 w_1 w_2$CDM$+A_L$ models.}
{\begin{tabular}{@{}lcccc@{}} \toprule
                                 & \multicolumn{2}{c}{Flat $w_0 w_2$CDM model}  &   \multicolumn{2}{c}{Flat $w_0 w_2$CDM$+A_L$ model} \\[+1mm]
\cline{2-3} \cline{4-5}\\[-1mm]
   Data                          &  P18 vs non-CMB  & P18+lensing vs non-CMB   &   P18 vs non-CMB  & P18+lensing vs non-CMB   \\
 \colrule
  $\log_{10} \mathcal{I}$        &   $-0.605$       &  $-0.496$                &   $0.202$         &  $0.186$     \\
  $\sigma$                       &   $2.538$        &  $2.462$                 &   $1.833$         &  $1.851$     \\
  $p$ (\%)                       &   $1.114$        &  $1.381$                 &   $6.685$         &  $6.412$     \\
\colrule
                                 & \multicolumn{2}{c}{Flat $w_0 w_p p$CDM model}  &   \multicolumn{2}{c}{Flat $w_0 w_p p$CDM$+A_L$ model} \\[+1mm]
\cline{2-3}  \cline{4-5}\\[-1mm]
   Data                          &  P18 vs non-CMB  & P18+lensing vs non-CMB   &   P18 vs non-CMB  & P18+lensing vs non-CMB   \\
 \colrule
  $\log_{10} \mathcal{I}$        &   $-0.411 $      &  $-0.308$                &   $0.344$         &  $0.358$     \\
  $\sigma$                       &   $2.461$        &  $2.315$                 &   $1.678$         &  $1.842$     \\
  $p$ (\%)                       &   $1.386 $       &  $2.064$                 &   $9.341$         &  $6.545$     \\
  \colrule
                                 & \multicolumn{2}{c}{Flat $w_0 w_1 w_2$CDM model}  &   \multicolumn{2}{c}{Flat $w_0 w_1 w_2$CDM$+A_L$ model} \\[+1mm]
\cline{2-3}  \cline{4-5}\\[-1mm]
   Data                          &  P18 vs non-CMB  & P18+lensing vs non-CMB   &   P18 vs non-CMB  & P18+lensing vs non-CMB   \\
 \colrule
  $\log_{10} \mathcal{I}$        &   $-0.609$       &  $-0.649$                &   $0.431$         &  $0.231$     \\
  $\sigma$                       &   $2.615$        &  $2.589$                 &   $1.872$         &  $1.901$     \\
  $p$ (\%)                       &   $0.894$        &  $0.964$                 &   $6.124$         &  $5.727$     \\
\botrule
\end{tabular}
\label{tab:consistency_w(z)CDM}}
\end{table}

To quantify how relatively well each model fits the dataset under consideration, we use differences in the Akaike information criterion ($\Delta$AIC) and deviance information criterion ($\Delta$DIC) between the information criterion (IC) values for the $w(z)$CDM parametrization under study and the $\Lambda$CDM model. See Sec.\ III of Ref.\ \refcite{deCruzPerez:2024abc} for a fuller discussion. According to the Jeffreys' scale we use, when $-2 \leq \Delta\textrm{IC}<0$ there is {\it weak} evidence in favor of the model under study, when $-6 \leq \Delta\textrm{IC} < -2$ there is {\it positive} evidence, when $-10\leq\Delta\textrm{IC} < -6$ there is {\it strong} evidence, and when $\Delta\textrm{IC} < -10$ there is {\it very strong} evidence in favor of the model under study relative to the tilted flat $\Lambda$CDM model. This scale also holds when $\Delta\textrm{IC}$ is positive, but then the flat $\Lambda$CDM model is favored over the $w(z)$CDM parametrization under study.

To quantitatively compare how consistent the cosmological parameter constraints (for the same parametrization) derived from two different data sets are, we use two estimators. The first is the DIC based $\log_{10}\mathcal{I}$, see Ref.\ \refcite{Joudaki:2016mvz} and Sec.\ III of Ref.\ \refcite{deCruzPerez:2024abc}. When the two data sets are consistent $\log_{10}\mathcal{I}>0$ while $\log_{10}\mathcal{I}<0$ means that the two data sets are inconsistent. According to the Jeffreys' scale we use, the degree of consistency or inconsistency between two data sets is {\it substantial} if $\lvert \log_{10}\mathcal{I} \rvert >0.5$, {\it strong} if $\lvert \log_{10}\mathcal{I} \rvert >1$, and {\it decisive} if $\lvert \log_{10}\mathcal{I} \rvert >2$ \cite{Joudaki:2016mvz}. The second estimator is the tension probability $p$ and the related, Gaussian approximation, "sigma value" $\sigma$, see Refs.\ \citen{Handley:2019pqx, Handley:2019wlz, Handley:2019tkm} and Sec.\ III of Ref.\ \refcite{deCruzPerez:2024abc}. $p=0.05$  and $p=0.003$ correspond to 2$\sigma$ and 3$\sigma$ Gaussian standard deviation, respectively.

\section{Results and Discussion}
\label{sec:ResultsandDiscussion}

Table \ref{tab:consistency_w(z)CDM} lists the values of the two statistical estimators, $\log_{10} \mathcal{I}$ and $p$ values, that measure the consistency between cosmological parameter constraints from P18 and non-CMB data and from P18+lensing and non-CMB data, for the three pairs of $w(z)$CDM and $w(z)$CDM$+A_L$ parametrizations we study in this paper. Compared to the corresponding values of the two statistical estimators in the $w_0w_a$CDM and $w_0w_a$CDM$+A_L$ cases \cite{Chan-GyungPark:2024mlx, Chan-GyungPark:2024brx}, the values here for all six parametrizations we study indicate better consistency between the pairs of dataset constraints. Patterns similar to those seen in Refs.\ \citen{Chan-GyungPark:2024mlx, Chan-GyungPark:2024brx} for the $w_0 w_a$CDM ($+A_L$) parametrizations are also found here. Allowing $A_L$ to vary increases the consistency between the pairs of dataset constraints, but even for the $A_L = 1$ cases the pairs of dataset constraints agree to better than 3$\sigma$. As these dataset constraints are consistent, in the following we mostly focus on results from the largest dataset, the P18+lensing+non-CMB one.


\begin{table}[htbp]
\tbl{Mean and 68\% (or 95\%) confidence limits of flat $w_0 w_2$CDM model parameters
        from non-CMB, P18, P18+lensing, P18+non-CMB, and P18+lensing+non-CMB data.
        $H_0$ has units of km s$^{-1}$ Mpc$^{-1}$. We also list the values of $\chi^2_{\text{min}}$, DIC, and AIC and the differences with respect to the values in the flat $\Lambda$CDM model for the same dataset, denoted by $\Delta\chi^2_{\text{min}}$, $\Delta$DIC, and $\Delta$AIC, respectively.}
{\begin{tabular}{@{}cccccc@{}} \toprule
  Parameter                     &  Non-CMB                     & P18                         &  P18+lensing             &  P18+non-CMB            & P18+lensing+non-CMB    \\
 \colrule
  $\Omega_b h^2$                & $0.0311^{+0.0039}_{-0.0045}$ & $0.02240 \pm 0.00015$       & $0.02243 \pm 0.00015$    &  $0.02243 \pm 0.00014$  &  $0.02244 \pm 0.00014$ \\
  $\Omega_c h^2$                & $0.1011^{+0.0065}_{-0.012}$  & $0.1199 \pm 0.0014$         & $0.1193 \pm 0.0012$      &  $0.1192 \pm 0.0011$    &  $0.11917 \pm 0.00099$ \\
  $100\theta_\textrm{MC}$       & $1.0234^{+0.0092}_{-0.011}$  & $1.04094 \pm 0.00031$       & $1.04100 \pm 0.00030$    &  $1.04098 \pm 0.00030$  &  $1.04098 \pm 0.00030$ \\
  $\tau$                        & $0.0540$                     & $0.0540 \pm 0.0079$         & $0.0520 \pm 0.0075$      &  $0.0523 \pm 0.0078$    &  $0.0527 \pm 0.0074$  \\
  $n_s$                         & $0.9655$                     & $0.9655 \pm 0.0043$         & $0.9668 \pm 0.0042$      &  $0.9668 \pm 0.0040$    &  $0.9668 \pm 0.0038$  \\
  $\ln(10^{10} A_s)$            & $3.54\pm 0.27$ ($>3.04$)     & $3.043 \pm 0.016$           & $3.037 \pm 0.015$        &  $3.038 \pm 0.016$      &  $3.039 \pm 0.014$   \\
  $w_0$                         & $-0.864 \pm 0.040$           & $-1.46^{+0.21}_{-0.44}$     & $-1.44^{+0.21}_{-0.45}$  &  $-0.898 \pm 0.041$     &  $-0.898 \pm 0.040$  \\  
  $w_2$                         & $-0.01^{+0.63}_{-0.25}$      & $-1.1 \pm 1.3$ ($< 1.24$)   & $-1.0 \pm 1.3$ ($<1.34$) &  $-1.12^{+0.51}_{-0.42}$&  $-1.12^{+0.50}_{-0.40}$\\
 \colrule
  $w_0+w_2$                     & $-0.88^{+0.62}_{-0.23}$      & $-2.56^{+0.90}_{-1.5}$      & $-2.5^{+1.0}_{-1.6}$     &  $-2.02^{+0.48}_{-0.39}$&  $-2.02^{+0.47}_{-0.37}$\\ 
  $H_0$                         & $69.8^{+2.2}_{-2.5}$         & $85 \pm 10$ ($> 66.0$)      & $85 \pm 10$ ($>66.7$)    &  $67.90 \pm 0.64$       &  $67.94 \pm 0.64$     \\
  $\Omega_m$                    & $0.2724^{+0.0095}_{-0.017}$  & $0.207^{+0.014}_{-0.065}$   & $0.208^{+0.015}_{-0.066}$&  $0.3087 \pm 0.0063$    &  $0.3083 \pm 0.0063$  \\
  $\sigma_8$                    & $0.819^{+0.031}_{-0.027}$    & $0.961^{+0.11}_{-0.045}$    & $0.951^{+0.11}_{-0.045}$ &  $0.812 \pm 0.011$      &  $0.8125 \pm 0.0091$  \\
      \colrule
  $\chi_{\textrm{min}}^2$       & $1457.50$                    & $2761.24$                   & $2770.45$                & $4233.13$               &  $4241.78$            \\
  $\Delta\chi_{\textrm{min}}^2$ & $-12.43$                     & $-4.56$                     & $-4.26$                  & $-7.11$                 &  $-7.48$              \\
  $\textrm{DIC}$                & $1470.99$                    & $2815.85$                   & $2824.55$                & $4289.62$               &  $4297.81$            \\
  $\Delta\textrm{DIC}$          & $-7.12$                      & $-2.08$                     & $-1.90$                  & $-2.71$                 &  $-3.39$             \\
  $\textrm{AIC}$                & $1469.50$                    & $2819.24$                   & $2828.45$                & $4291.13$               &  $4299.78$           \\
  $\Delta\textrm{AIC}$          & $-8.43$                      & $-0.56$                     & $-0.26$                  & $-3.11$                 &  $-3.48$            \\
\botrule
\end{tabular}\label{tab:results_flat_w0w2CDM}}
\end{table}



\begin{table}[htbp]
\tbl{Mean and 68\% (or 95\%) confidence limits of flat $w_0 w_2$CDM$+A_L$ model parameters from non-CMB, P18, P18+lensing, P18+non-CMB, and P18+lensing+non-CMB data. $H_0$ has units of km s$^{-1}$ Mpc$^{-1}$. We also list the values of $\chi^2_{\text{min}}$, DIC, and AIC and the differences with respect to the values in the flat $\Lambda$CDM model for the same dataset, denoted by $\Delta\chi^2_{\text{min}}$, $\Delta$DIC, and $\Delta$AIC, respectively.}
{\begin{tabular}{@{}cccccc@{}} \toprule
  Parameter                     &  Non-CMB                     & P18                         &  P18+lensing                 &  P18+non-CMB                 & P18+lensing+non-CMB    \\
 \colrule 
  $\Omega_b h^2$                & $0.0311^{+0.0039}_{-0.0045}$ & $0.02258 \pm 0.00017$       & $0.02250 \pm 0.00017$        &  $0.02263 \pm 0.00016$       &  $0.02255 \pm 0.00015$ \\
  $\Omega_c h^2$                & $0.1011^{+0.0065}_{-0.012}$  & $0.1182 \pm 0.0015$         & $0.1184 \pm 0.0015$          &  $0.1177 \pm 0.0012$         &  $0.1179 \pm 0.0012$ \\
  $100\theta_\textrm{MC}$       & $1.0234^{+0.0092}_{-0.011}$  & $1.04114 \pm 0.00033$       & $1.04108 \pm 0.00032$        &  $1.04119 \pm 0.00032$       &  $1.04114 \pm 0.00031$ \\
  $\tau$                        & $0.0540$                     & $0.0494 \pm 0.0087$         & $0.0495^{+0.0085}_{-0.0073}$ &  $0.0481^{+0.0089}_{-0.0074}$&  $0.0480 \pm 0.0083$  \\
  $n_s$                         & $0.9654$                     & $0.9706 \pm 0.0049$         & $0.9690 \pm 0.0048$          &  $0.9718 \pm 0.0044$         &  $0.9706 \pm 0.0042$  \\
  $\ln(10^{10} A_s)$            & $3.54\pm 0.27$ ($>3.04$)     & $3.030 \pm 0.018$           & $3.030^{+0.018}_{-0.016}$    &  $3.026^{+0.018}_{-0.015}$   &  $3.025 \pm 0.017$   \\
  $A_L$                         & \ldots                       & $1.164^{+0.064}_{-0.087}$   & $1.046^{+0.040}_{-0.055}$    &  $1.186 \pm 0.066$           &  $1.072 \pm 0.038$   \\
  $w_0$                         & $-0.864 \pm 0.040$           & $-1.15^{+0.35}_{-0.65}$     & $-1.28^{+0.27}_{-0.58}$      &  $-0.909 \pm 0.040$          &  $-0.908 \pm 0.040$  \\  
  $w_2$                         & $-0.01^{+0.63}_{-0.25}$      & $-1.0 \pm 1.3$ ($< 1.24$)   & $-0.9 \pm 1.3$ ($<1.35$)     &  $-0.73^{+0.48}_{-0.38}$     &  $-0.78^{+0.48}_{-0.39}$\\ 
 \colrule 
  $w_0+w_2$                     & $-0.88^{+0.62}_{-0.23}$      & $-2.1^{+1.4}_{-1.2}$        & $-2.2 \pm 1.2$               &  $-1.64^{+0.45}_{-0.35}$     &  $-1.69^{+0.45}_{-0.36}$\\ 
  $H_0$                         & $69.8^{+2.2}_{-2.5}$         & $77^{+20}_{-8}$ ($> 53.4$)  & $80 \pm 13$ ($>57.6$)        &  $67.96 \pm 0.63$            &  $67.93 \pm 0.63$     \\
  $\Omega_m$                    & $0.2724^{+0.0095}_{-0.017}$  & $0.268^{+0.038}_{-0.013}$   & $0.240^{+0.022}_{-0.099}$    &  $0.3052 \pm 0.0062$         &  $0.3058 \pm 0.0062$  \\
  $\sigma_8$                    & $0.819^{+0.031}_{-0.027}$    & $0.868^{+0.16}_{-0.085}$    & $0.901^{+0.15}_{-0.066}$     &  $0.795 \pm 0.012$           &  $0.797 \pm 0.012$  \\
 \colrule
  $\chi_{\textrm{min}}^2$       & $1457.50$                    & $2755.88$                   & $2770.27$                    & $4222.49$                    &  $4237.66$            \\
  $\Delta\chi_{\textrm{min}}^2$ & $-12.43$                     & $-9.92$                     & $-4.44$                      & $-17.75$                     &  $-11.60$              \\
  $\textrm{DIC}$                & $1470.99$                    & $2813.08$                   & $2825.76$                    & $4283.13$                    &  $4295.89$            \\
  $\Delta\textrm{DIC}$          & $-7.12$                      & $-4.85$                     & $-0.69$                      & $-9.20$                      &  $-5.31$             \\
  $\textrm{AIC}$                & $1469.50$                    & $2815.88$                   & $2830.27$                    & $4282.49$                    &  $4297.66$           \\
  $\Delta\textrm{AIC}$          & $-8.43$                      & $-3.92$                     & $+1.56$                      & $-11.75$                     &  $-5.60$            \\
\botrule
\end{tabular}\label{tab:results_flat_w0w2CDM_Alens}}
\end{table}

\begin{figure*}[htbp]
\centering
\mbox{\includegraphics[width=127mm]{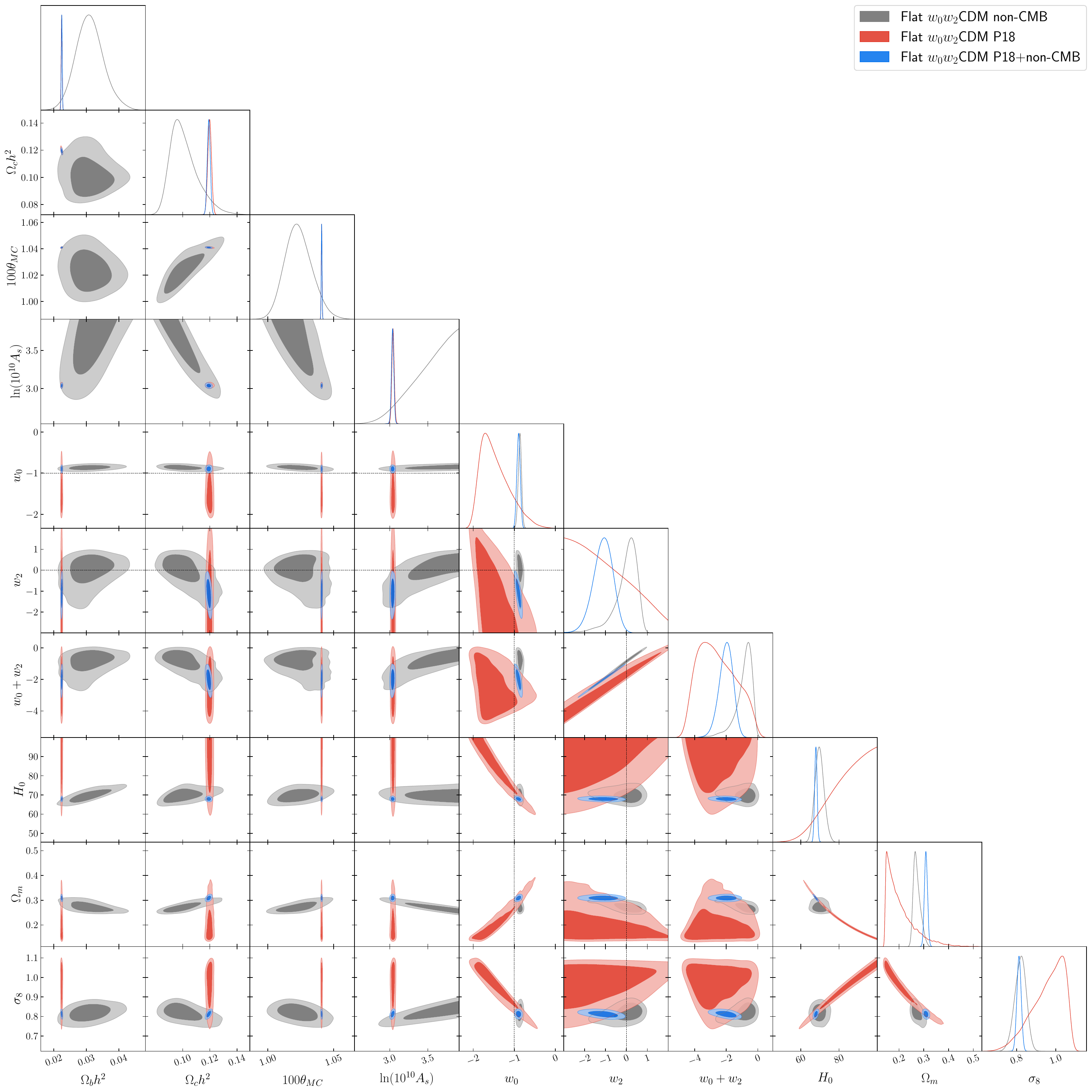}}
        \caption{One-dimensional likelihoods and 1$\sigma$ and $2\sigma$ likelihood confidence contours of flat $w_0 w_2$CDM model parameters favored by non-CMB, P18, and P18+non-CMB datasets. We do not show $\tau$ and $n_s$, which are fixed in the non-CMB data analysis.
}
\label{fig:flat_w0w2CDM_P18_nonCMB23v2}
\end{figure*}

\begin{figure*}[htbp]
\centering
\mbox{\includegraphics[width=127mm]{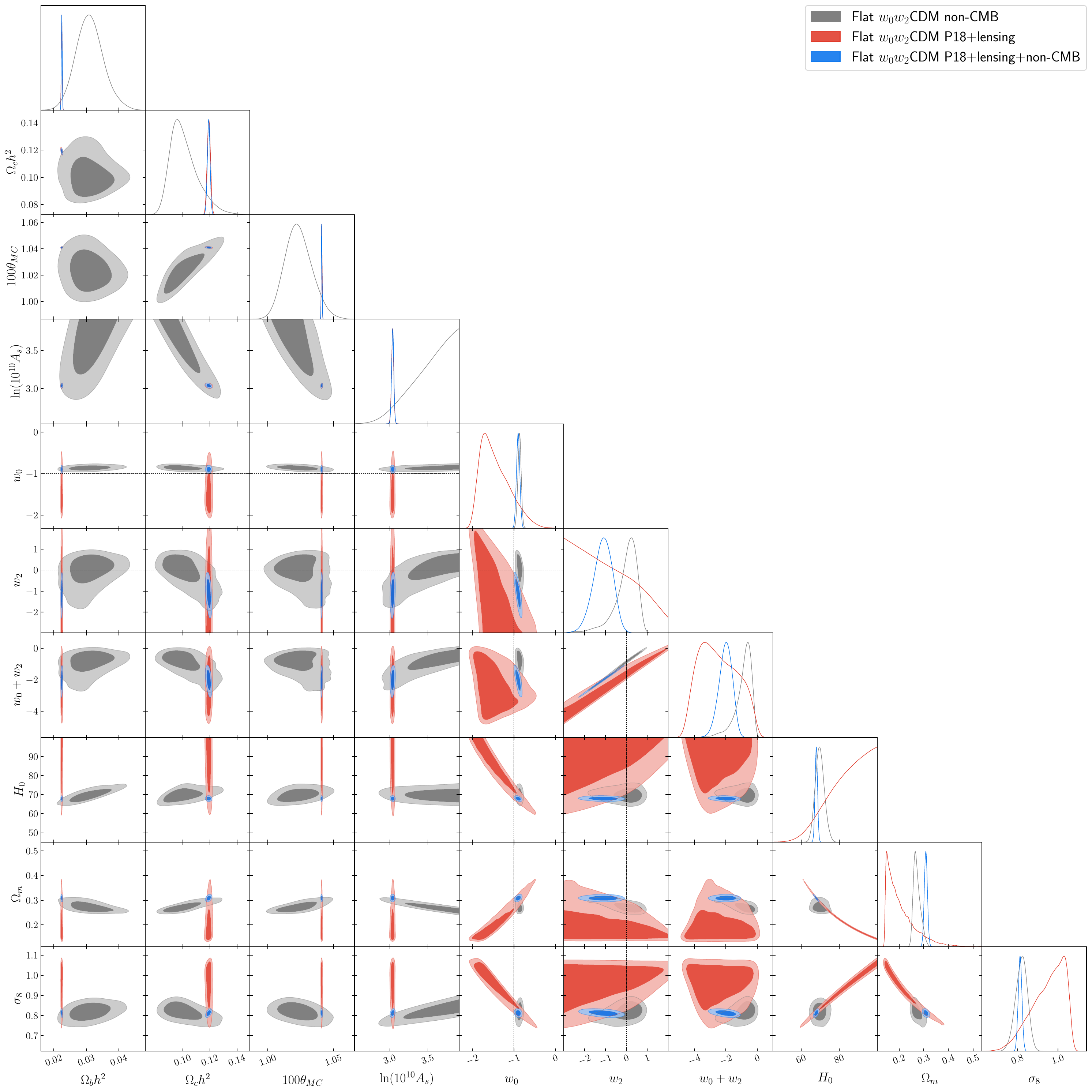}}
\caption{One-dimensional likelihoods and 1$\sigma$ and $2\sigma$ likelihood confidence contours of flat $w_0 w_2$CDM model parameters favored by non-CMB, P18+lensing, P18+lensing+non-CMB datasets. We do not show $\tau$ and $n_s$, which are fixed in the non-CMB data analysis.
}
\label{fig:flat_w0w2CDM_P18_lensing_nonCMB23v2}
\end{figure*}


\begin{figure*}[htbp]
\centering
\mbox{\includegraphics[width=127mm]{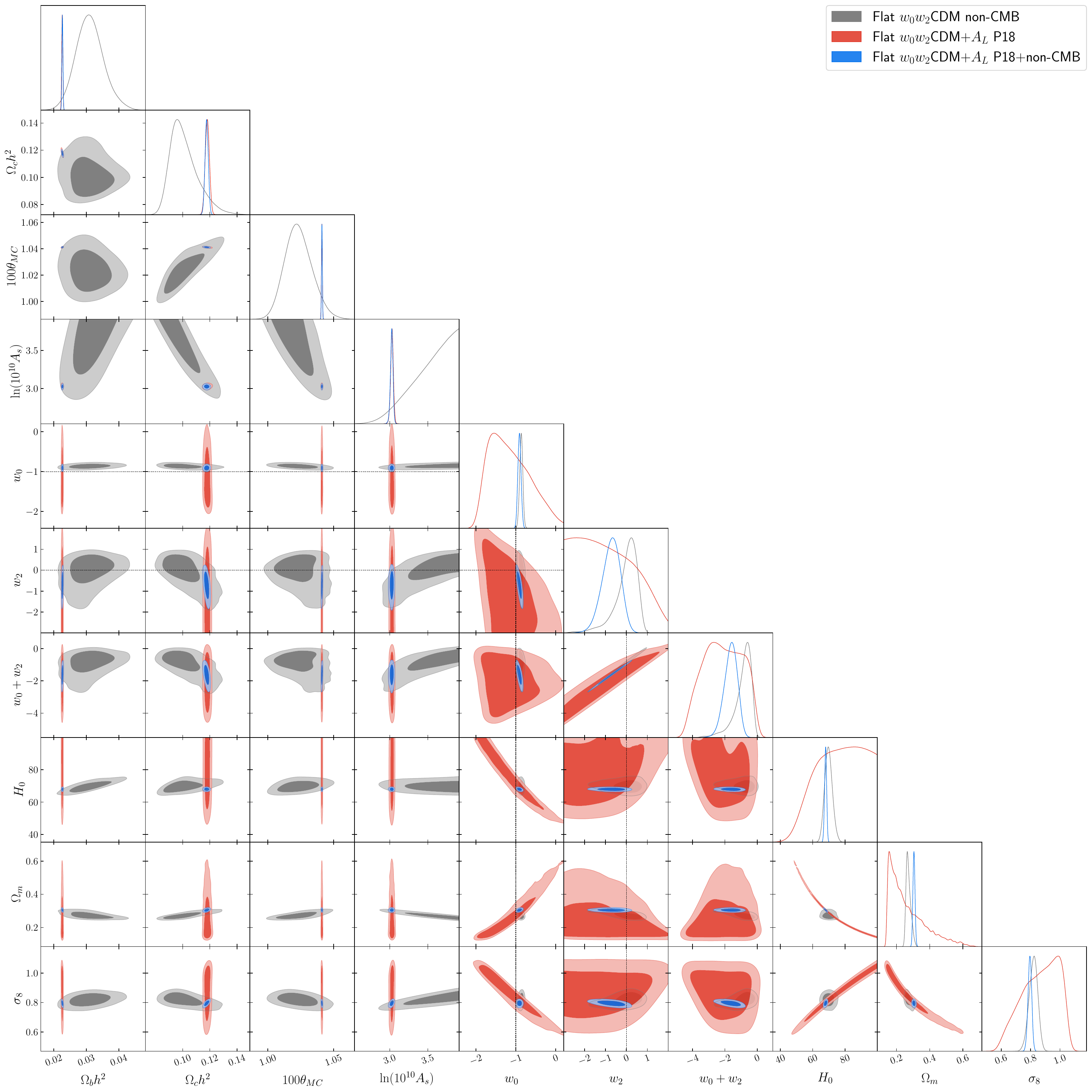}}
        \caption{One-dimensional likelihoods and 1$\sigma$ and $2\sigma$ likelihood confidence contours of flat $w_0 w_2$CDM$+A_L$ model parameters favored by non-CMB, P18, and P18+non-CMB datasets. We do not show $\tau$ and $n_s$, which are fixed in the non-CMB data analysis.
}
\label{fig:flat_w0w2CDM_Alens_P18_nonCMB23v2}
\end{figure*}

\begin{figure*}[htbp]
\centering
\mbox{\includegraphics[width=127mm]{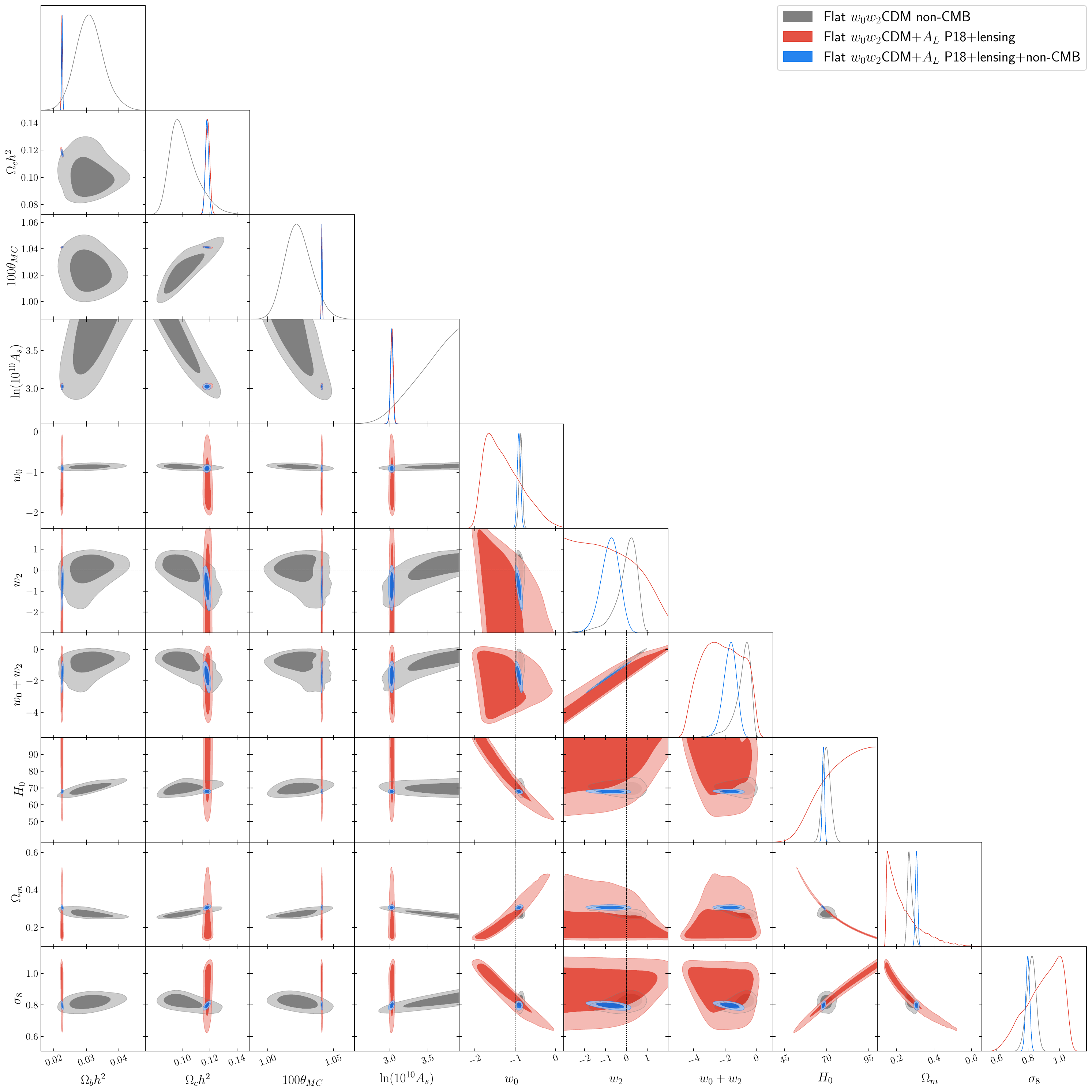}}
\caption{One-dimensional likelihoods and 1$\sigma$ and $2\sigma$ likelihood confidence contours of flat $w_0 w_2$CDM$+A_L$ model parameters favored by non-CMB, P18+lensing, P18+lensing+non-CMB datasets. We do not show $\tau$ and $n_s$, which are fixed in the non-CMB data analysis.
}
\label{fig:flat_w0w2CDM_Alens_P18_lensing_nonCMB23v2}
\end{figure*}


\begin{figure*}[htbp]
\centering
\mbox{\includegraphics[width=62mm]{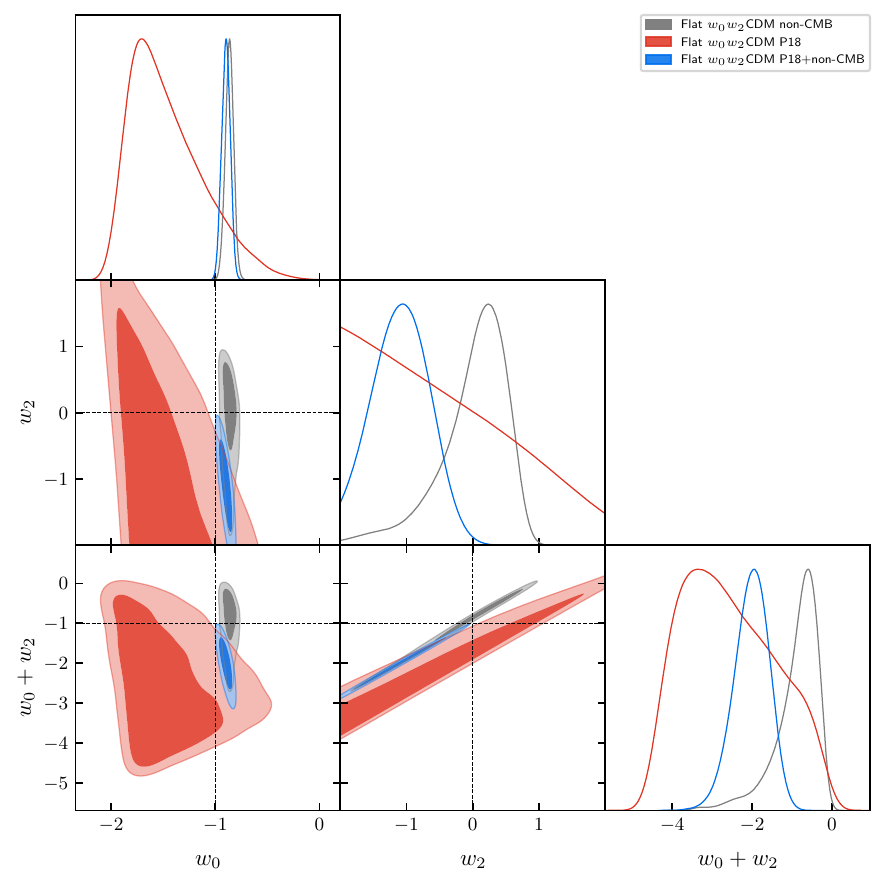}}
\mbox{\includegraphics[width=62mm]{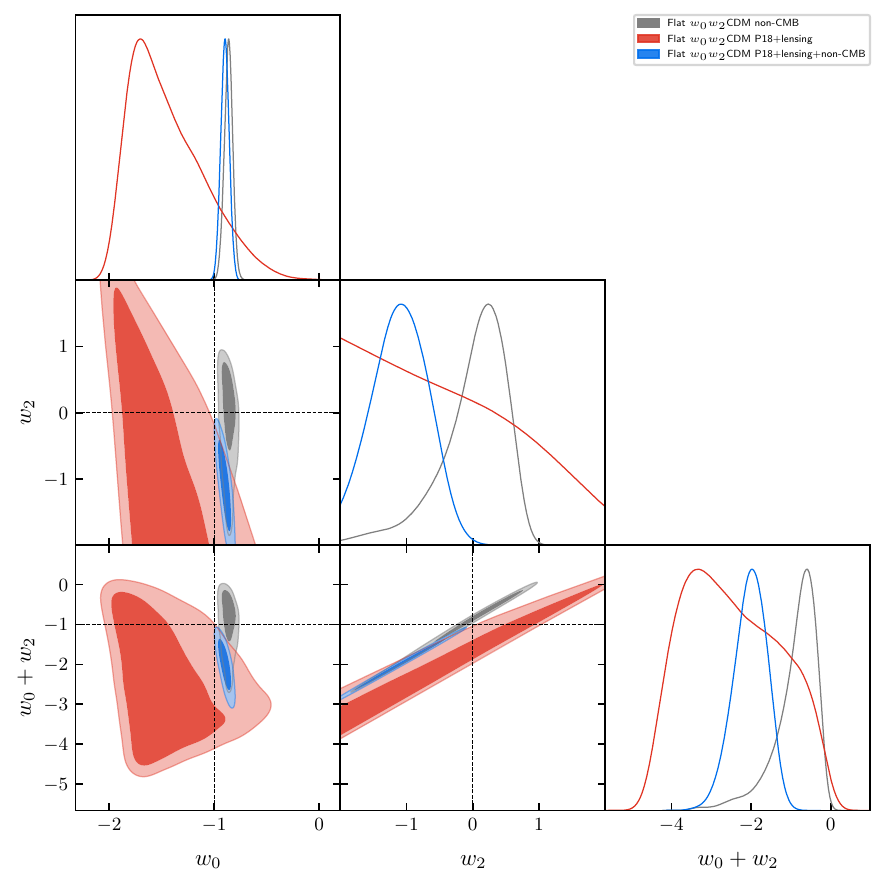}}
        \caption{One-dimensional likelihoods and 1$\sigma$ and $2\sigma$ likelihood confidence contours of $w_0$, $w_2$, and $w_0+w_2$ parameters in the flat $w_0 w_2$CDM parametrization favored by (left) non-CMB, P18, and P18+non-CMB datasets, and (right) non-CMB, P18+lensing, and P18+lensing+non-CMB datasets.
}
\label{fig:flat_w0w2CDM_P18_nonCMB23v2_w0w2}
\end{figure*}


\begin{figure*}[htbp]
\centering
\mbox{\includegraphics[width=62mm]{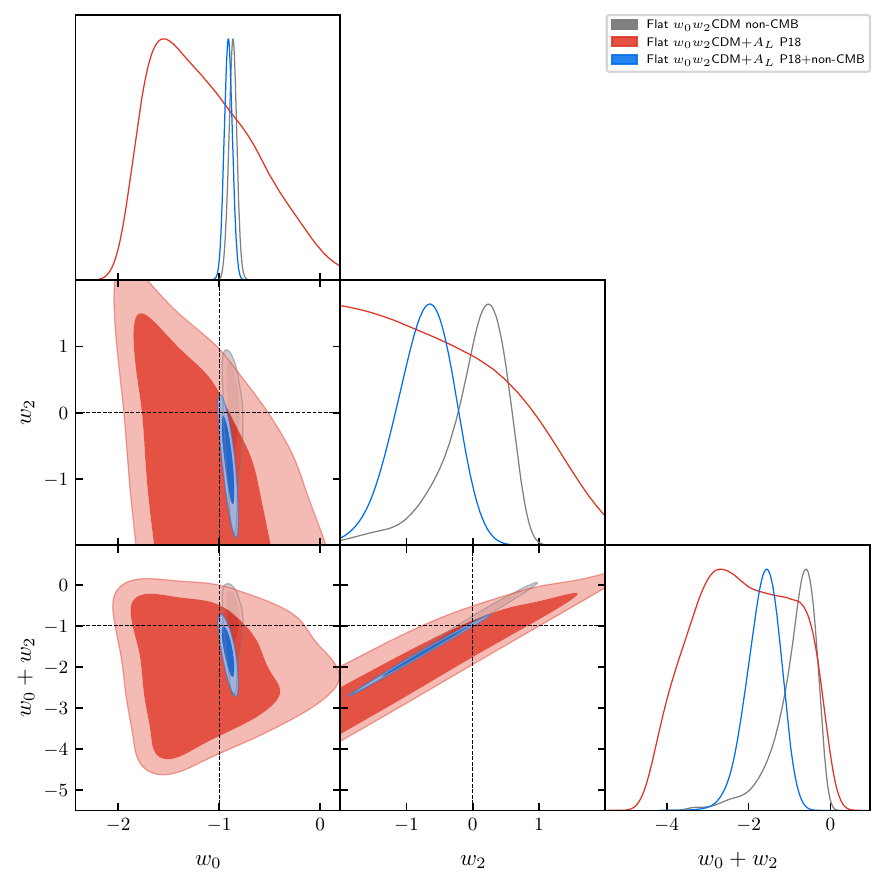}}
\mbox{\includegraphics[width=62mm]{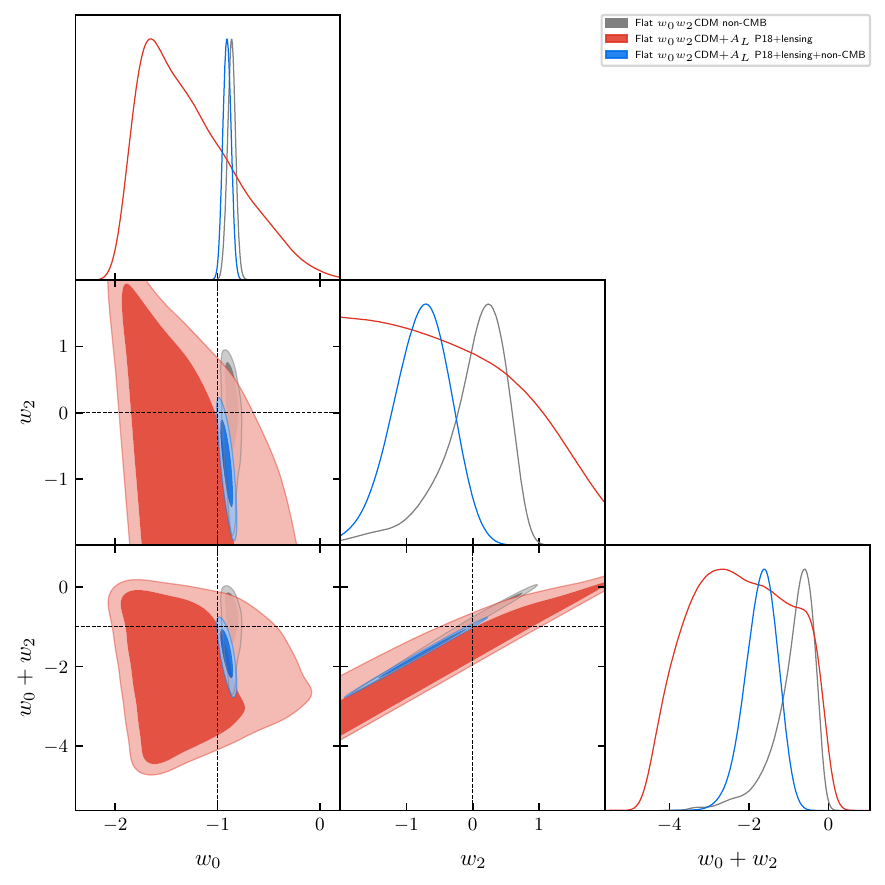}}
        \caption{One-dimensional likelihoods and 1$\sigma$ and $2\sigma$ likelihood confidence contours of $w_0$, $w_2$, and $w_0+w_2$ parameters in the flat $w_0 w_2$CDM$+A_L$ parametrization favored by (left) non-CMB, P18, and P18+non-CMB datasets, and (right) non-CMB, P18+lensing, and P18+lensing+non-CMB datasets.
}
\label{fig:flat_w0w2CDM_Alens_P18_nonCMB23v2_w0w2}
\end{figure*}

\begin{figure*}[htbp]
\centering
\mbox{\includegraphics[width=127mm]{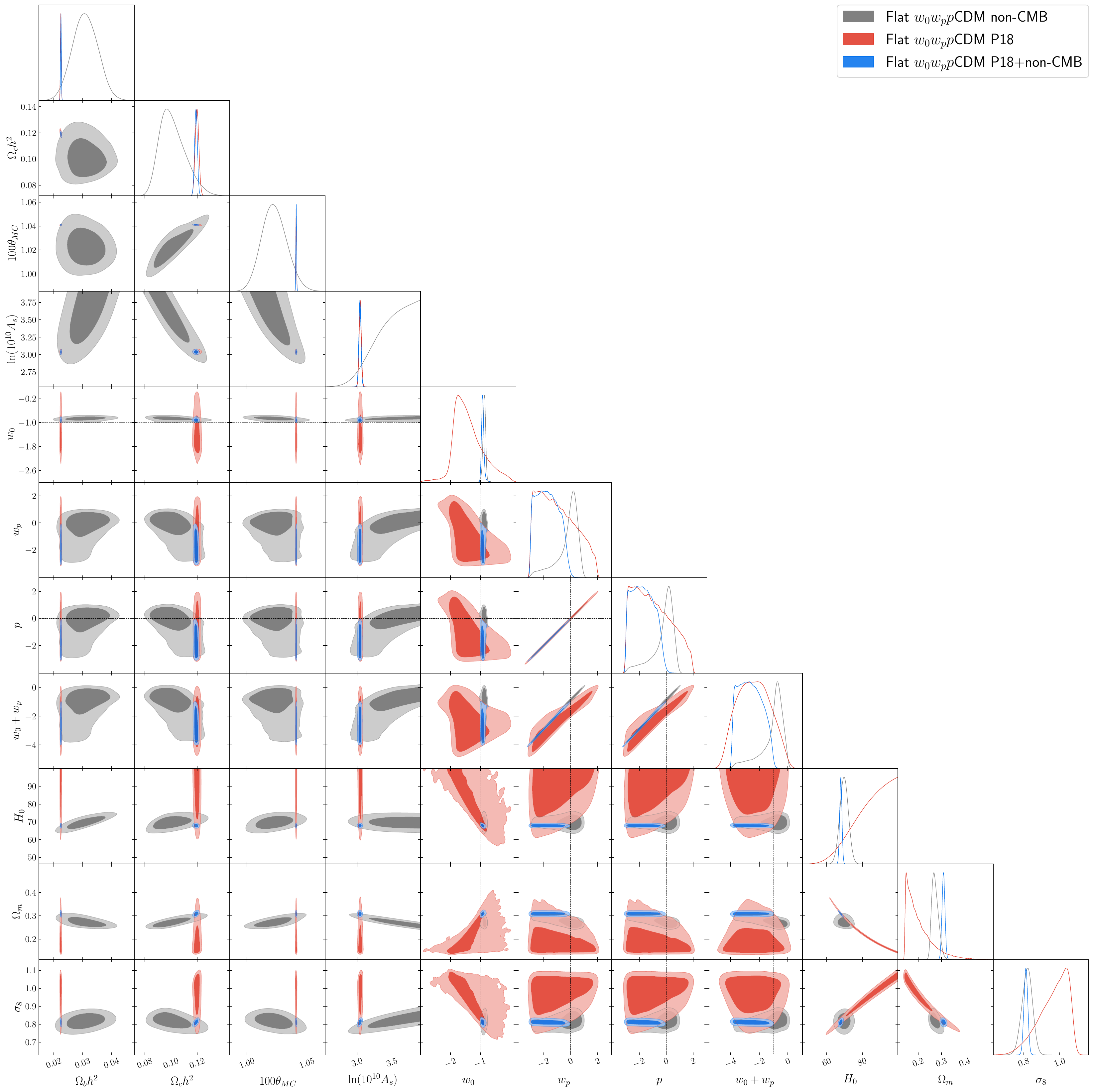}}
        \caption{One-dimensional likelihoods and 1$\sigma$ and $2\sigma$ likelihood confidence contours of flat $w_0 w_p p$CDM model parameters favored by non-CMB, P18, and P18+non-CMB datasets. We do not show $\tau$ and $n_s$, which are fixed in the non-CMB data analysis.
}
\label{fig:flat_w0w2p2CDM_P18_nonCMB23v2}
\end{figure*}
\begin{figure*}[htbp]
\centering
\mbox{\includegraphics[width=127mm]{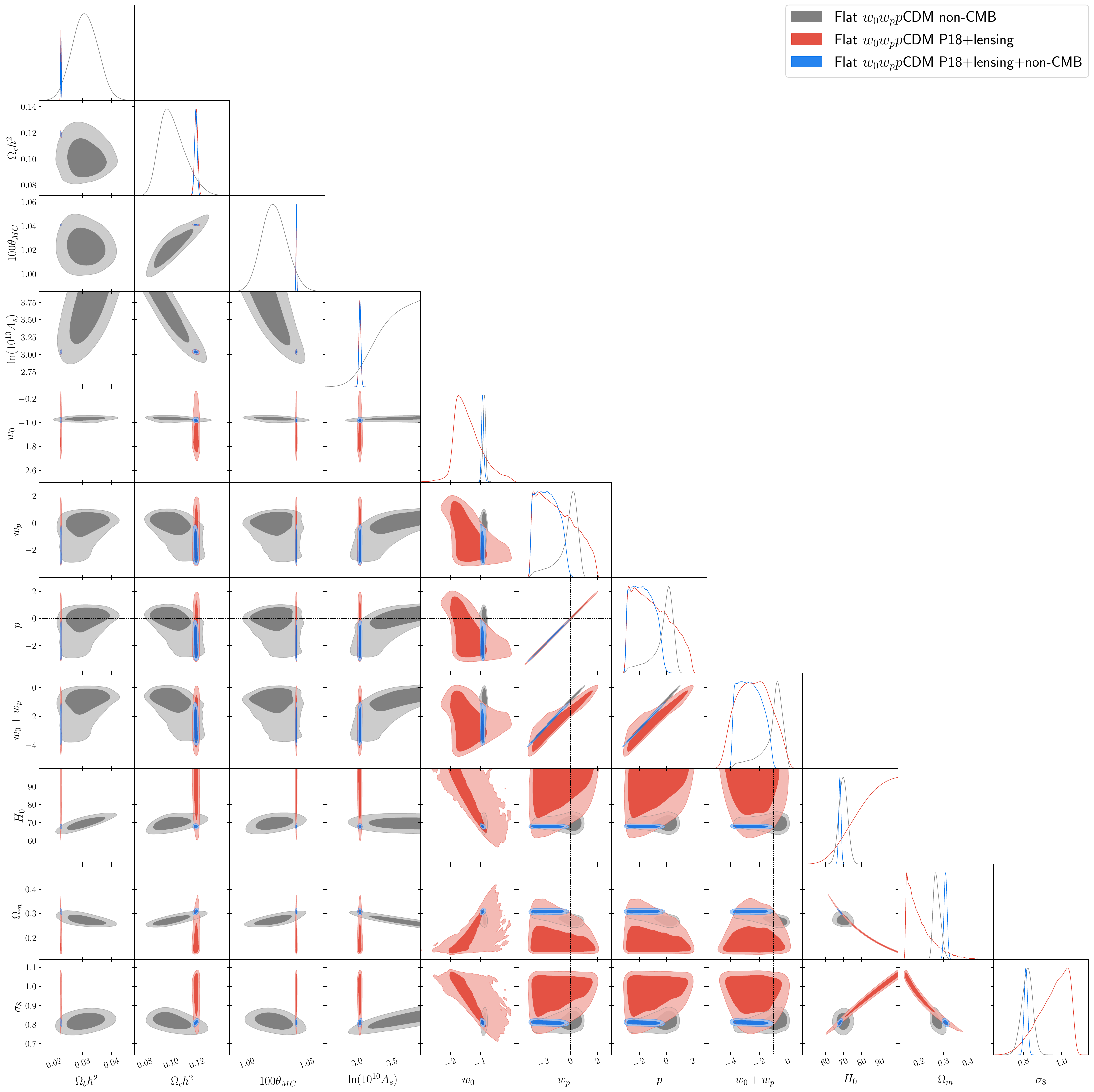}}
\caption{One-dimensional likelihoods and 1$\sigma$ and $2\sigma$ likelihood confidence contours of flat $w_0 w_p p$CDM model parameters favored by non-CMB, P18+lensing, P18+lensing+non-CMB datasets. We do not show $\tau$ and $n_s$, which are fixed in the non-CMB data analysis.
}
\label{fig:flat_w0w2p2CDM_P18_lensing_nonCMB23v2}
\end{figure*}
\begin{figure*}[htbp]
\centering
\mbox{\includegraphics[width=127mm]{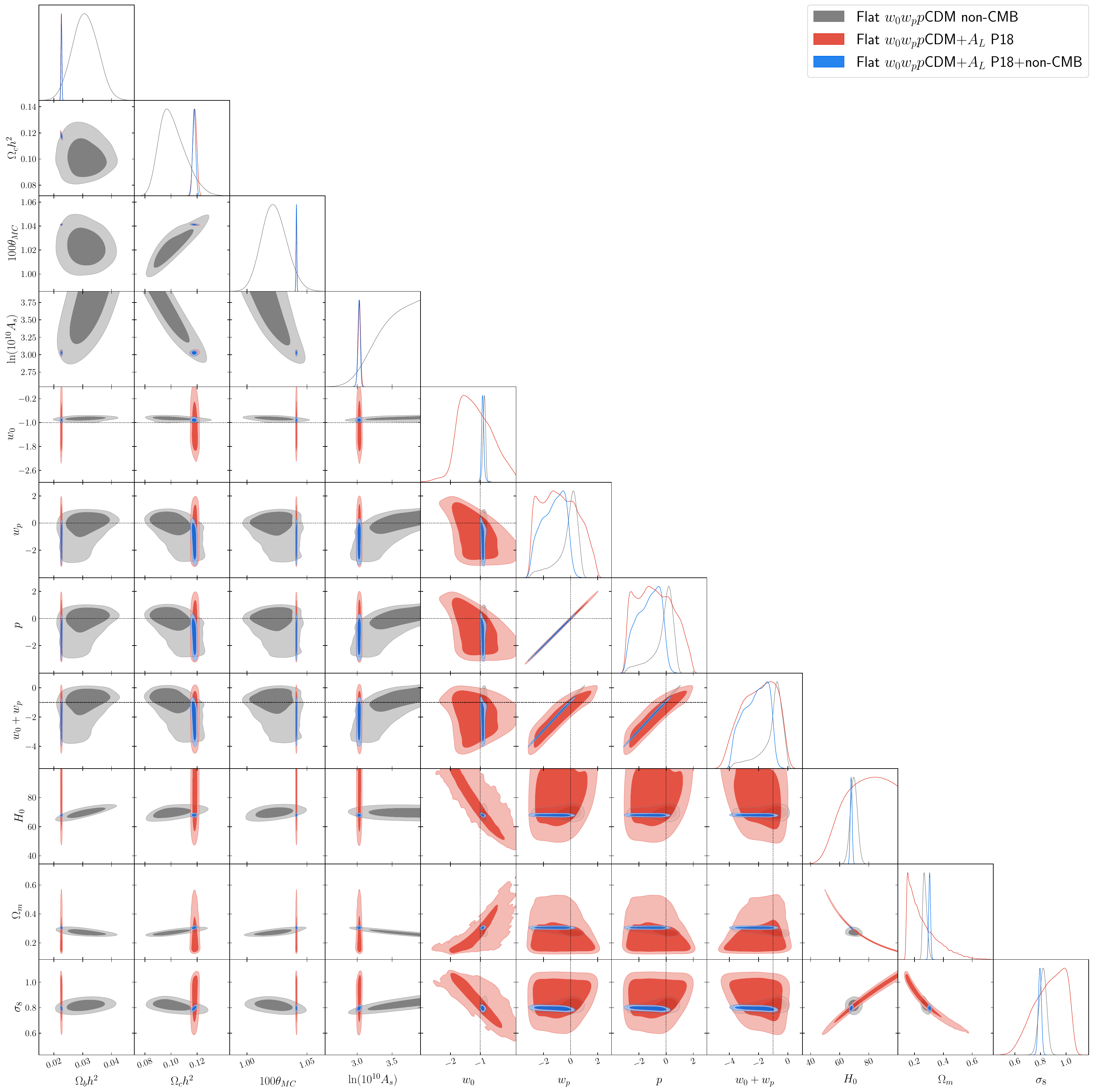}}
        \caption{One-dimensional likelihoods and 1$\sigma$ and $2\sigma$ likelihood confidence contours of flat $w_0 w_p p$CDM$+A_L$ model parameters favored by non-CMB, P18, and P18+non-CMB datasets. We do not show $\tau$ and $n_s$, which are fixed in the non-CMB data analysis.
}
\label{fig:flat_w0w2p2CDM_Alens_P18_nonCMB23v2}
\end{figure*}
\begin{figure*}[htbp]
\centering
\mbox{\includegraphics[width=127mm]{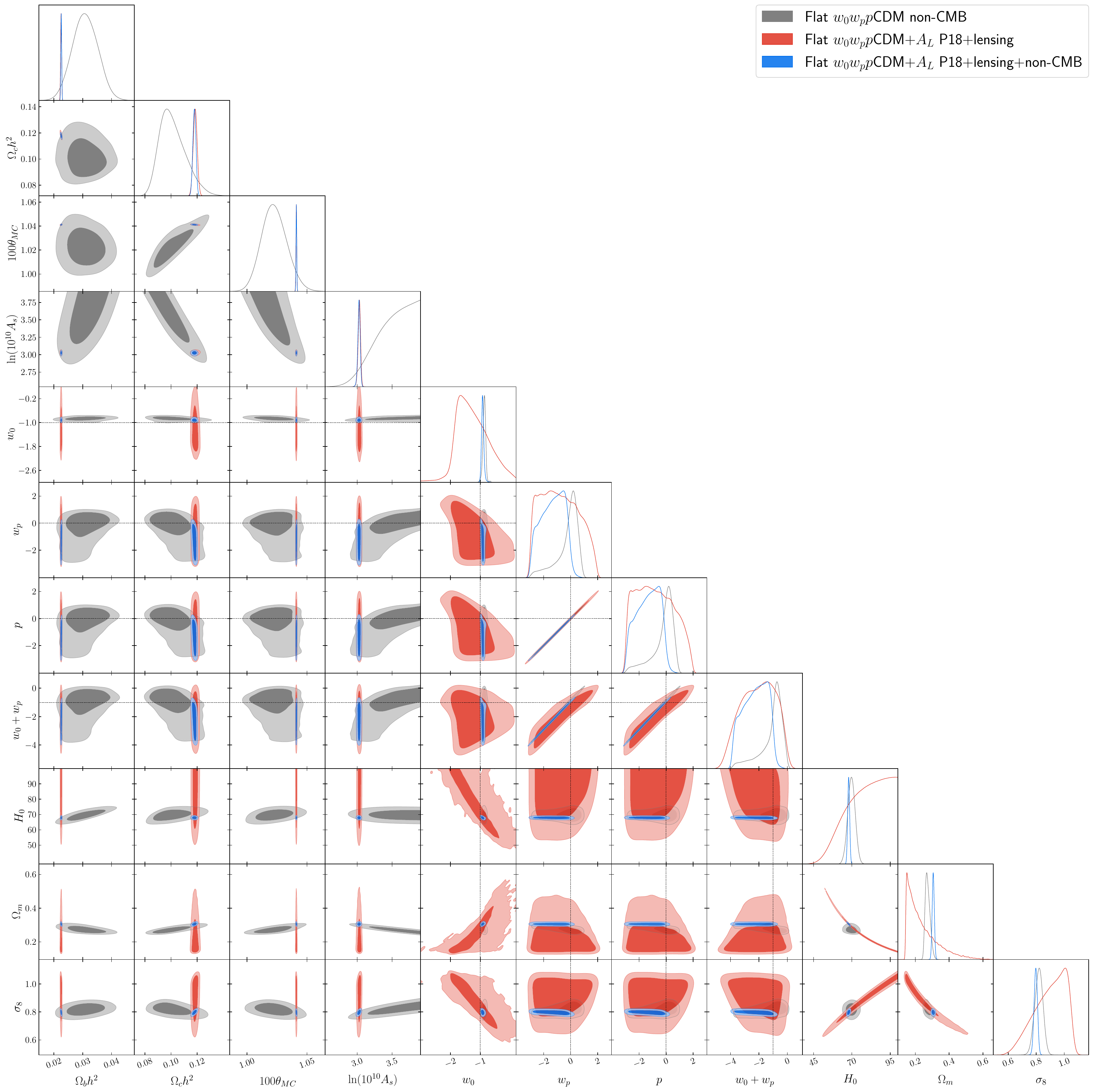}}
\caption{One-dimensional likelihoods and 1$\sigma$ and $2\sigma$ likelihood confidence contours of flat $w_0 w_p p$CDM$+A_L$ model parameters favored by non-CMB, P18+lensing, P18+lensing+non-CMB datasets. We do not show $\tau$ and $n_s$, which are fixed in the non-CMB data analysis.
}
\label{fig:flat_w0w2p2CDM_Alens_P18_lensing_nonCMB23v2}
\end{figure*}
\begin{figure*}[htbp]
\centering
\mbox{\includegraphics[width=62mm]{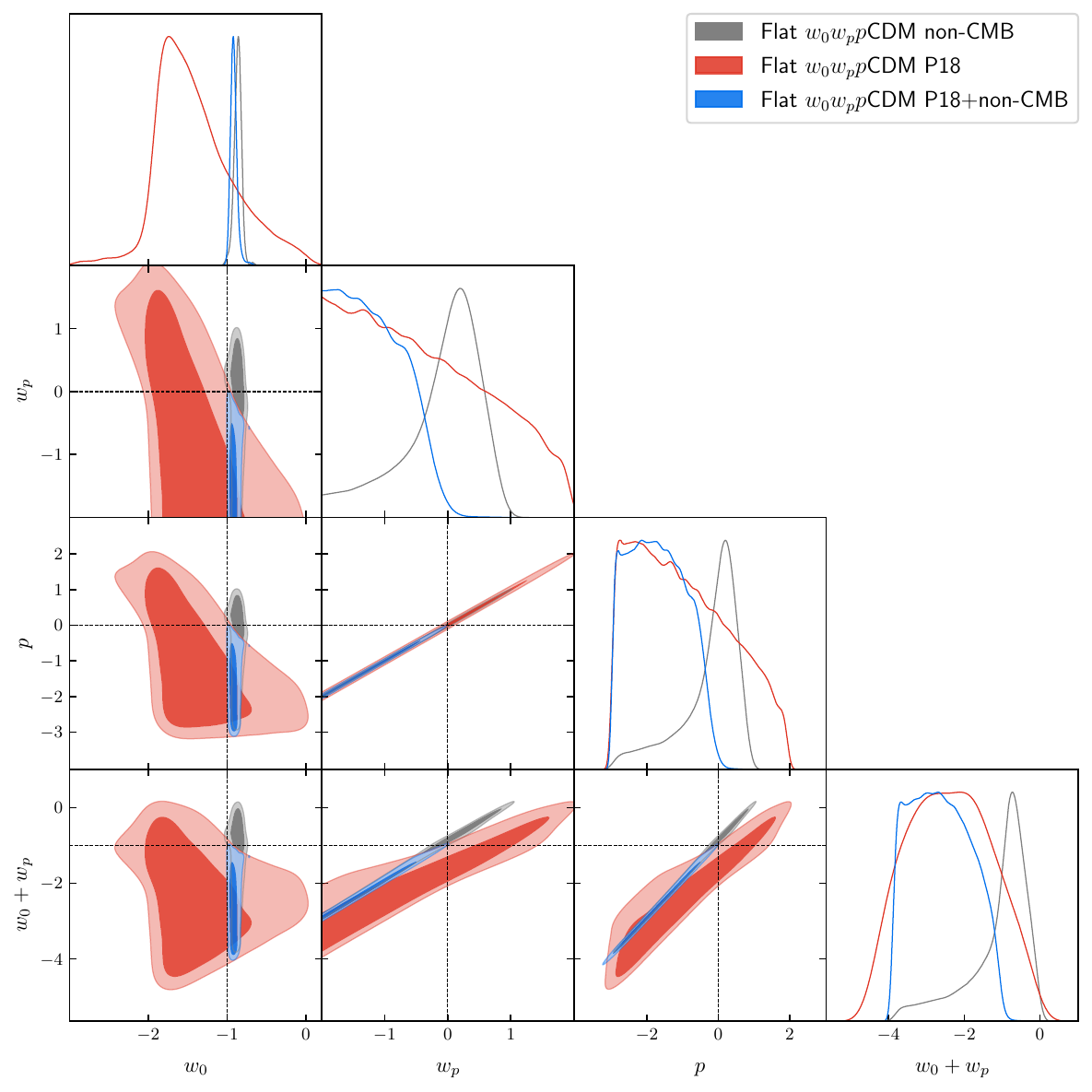}}
\mbox{\includegraphics[width=62mm]{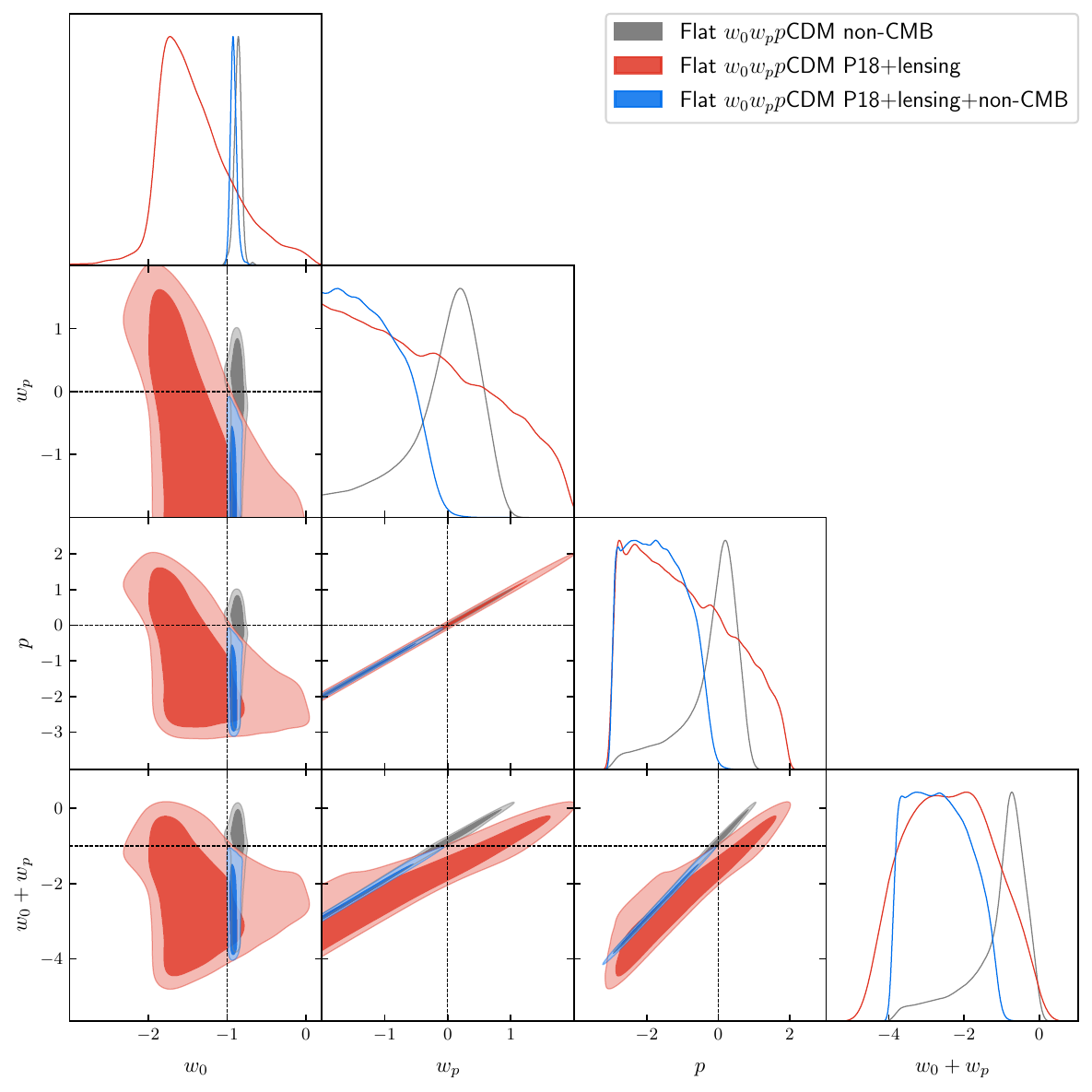}}
        \caption{One-dimensional likelihoods and 1$\sigma$ and $2\sigma$ likelihood confidence contours of $w_0$, $w_p$, $p$, and $w_0+w_p$ parameters in the flat $w_0 w_p p$CDM parametrization favored by (left) non-CMB, P18, and P18+non-CMB datasets, and (right) non-CMB, P18+lensing, and P18+lensing+non-CMB datasets.
}
\label{fig:flat_w0w2p2CDM_P18_nonCMB23v2_w0w2p2}
\end{figure*}
\begin{figure*}[htbp]
\centering
\mbox{\includegraphics[width=62mm]{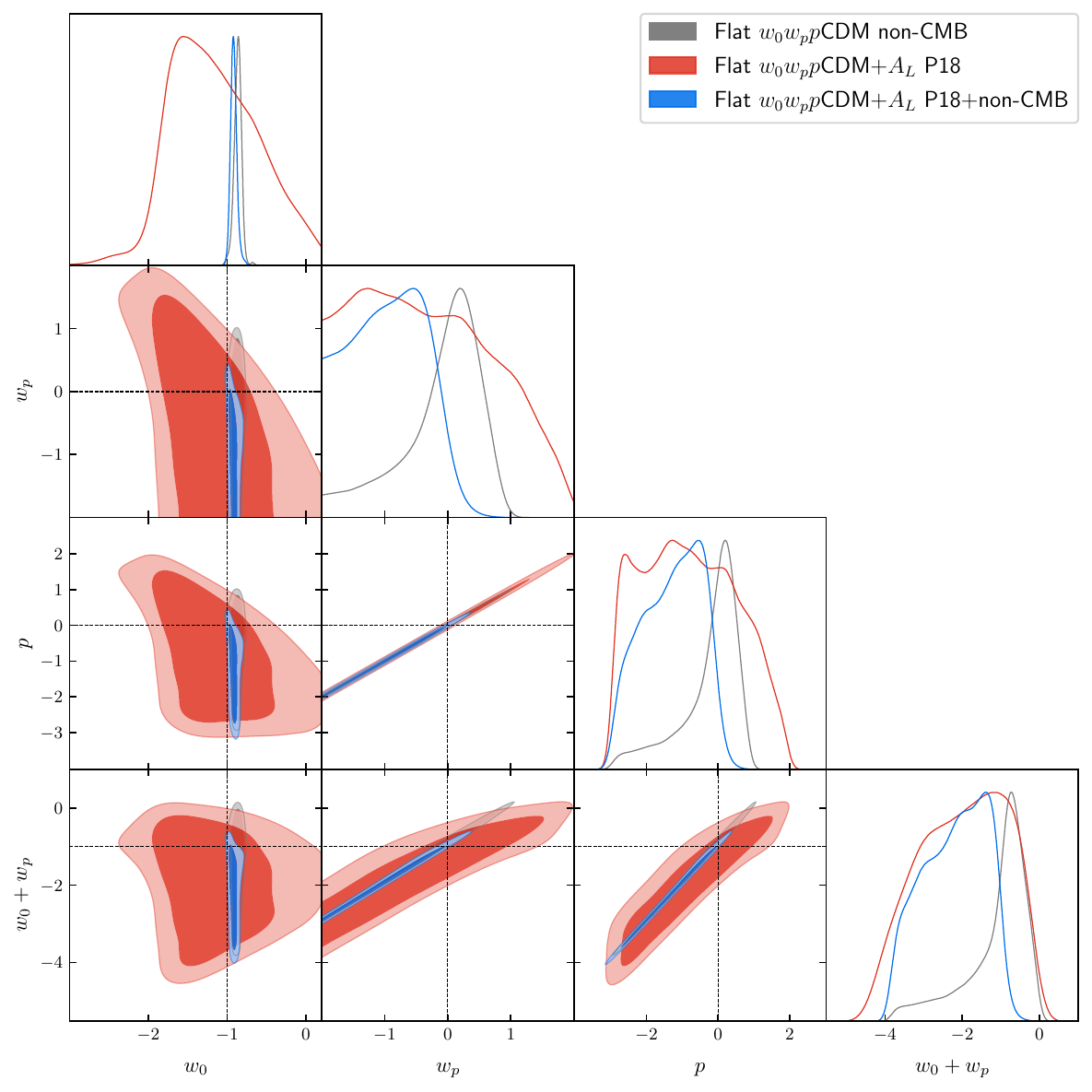}}
\mbox{\includegraphics[width=62mm]{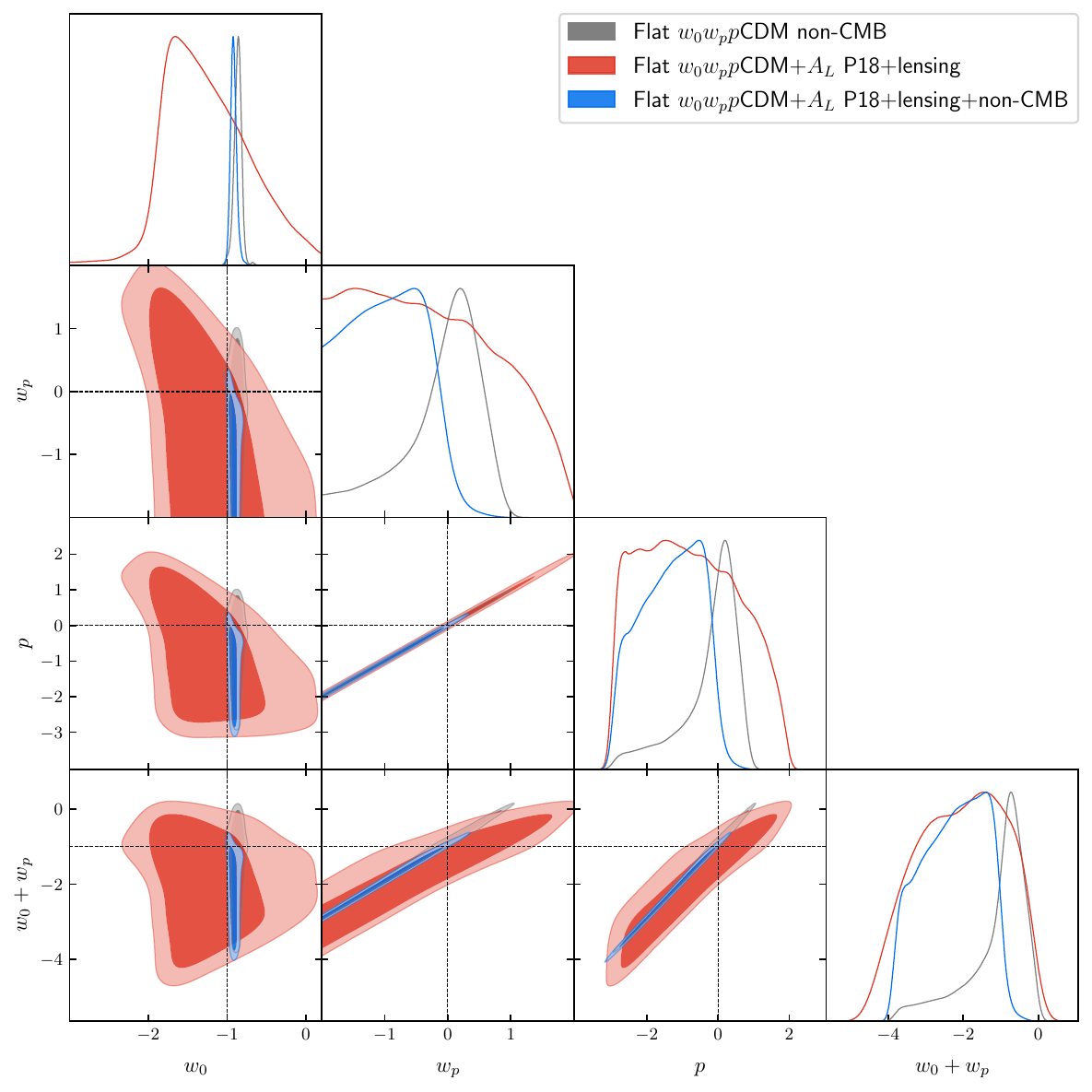}}
        \caption{One-dimensional likelihoods and 1$\sigma$ and $2\sigma$ likelihood confidence contours of $w_0$, $w_p$, $p$, and $w_0+w_p$ parameters in the flat $w_0 w_p p$CDM$+A_L$ model favored by (left) non-CMB, P18, and P18+non-CMB datasets, and (right) non-CMB, P18+lensing, and P18+lensing+non-CMB datasets.
}
\label{fig:flat_w0w2p2CDM_Alens_P18_nonCMB23v2_w0w2p2}
\end{figure*}
\begin{figure*}[htbp]
\centering
\mbox{\includegraphics[width=127mm]{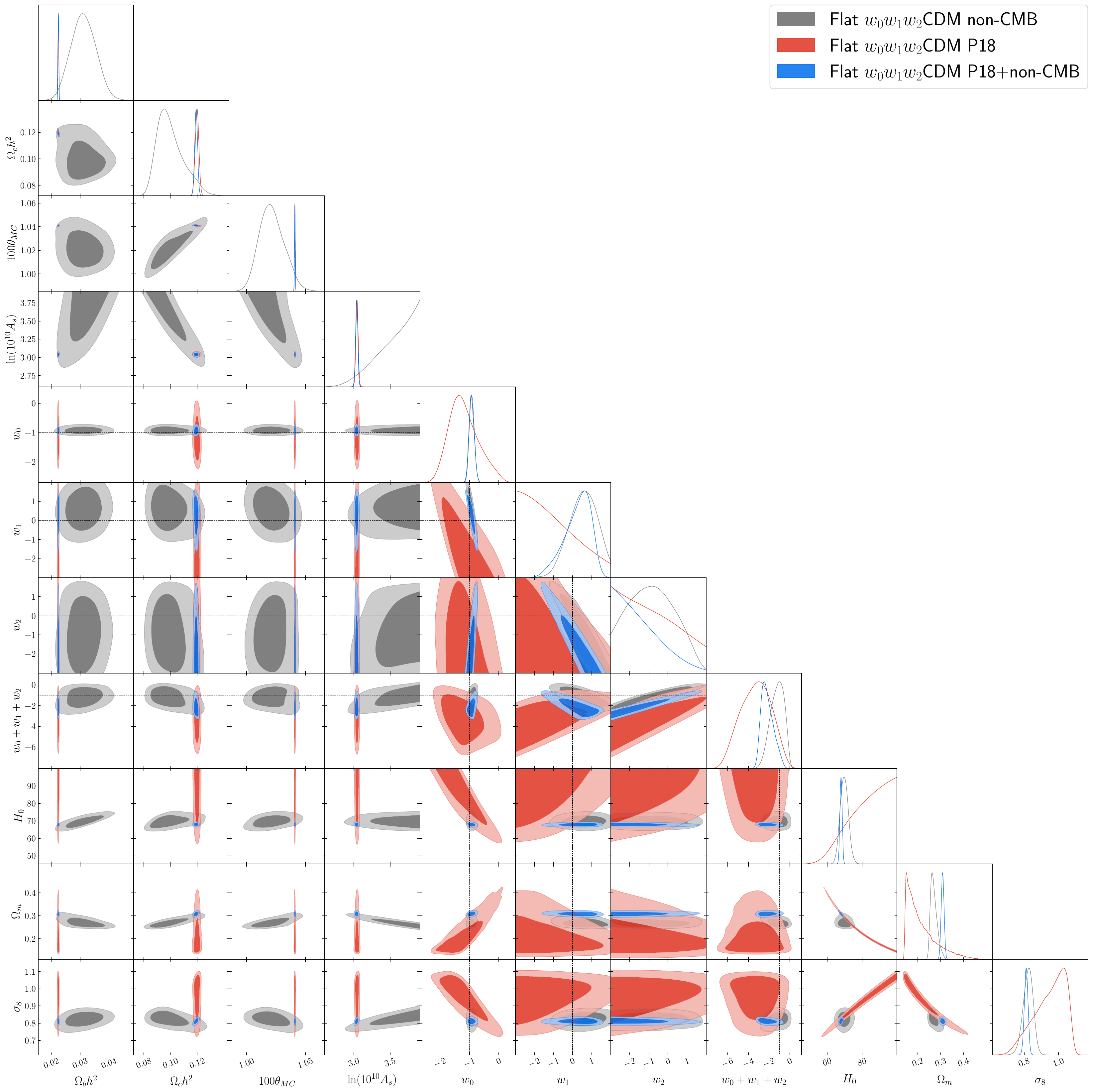}}
        \caption{One-dimensional likelihoods and 1$\sigma$ and $2\sigma$ likelihood confidence contours of flat $w_0 w_1 w_2$CDM model parameters favored by non-CMB, P18, and P18+non-CMB datasets. We do not show $\tau$ and $n_s$, which are fixed in the non-CMB data analysis.
}
\label{fig:flat_w0w1w2CDM_P18_nonCMB23v2}
\end{figure*}
\begin{figure*}[htbp]
\centering
\mbox{\includegraphics[width=127mm]{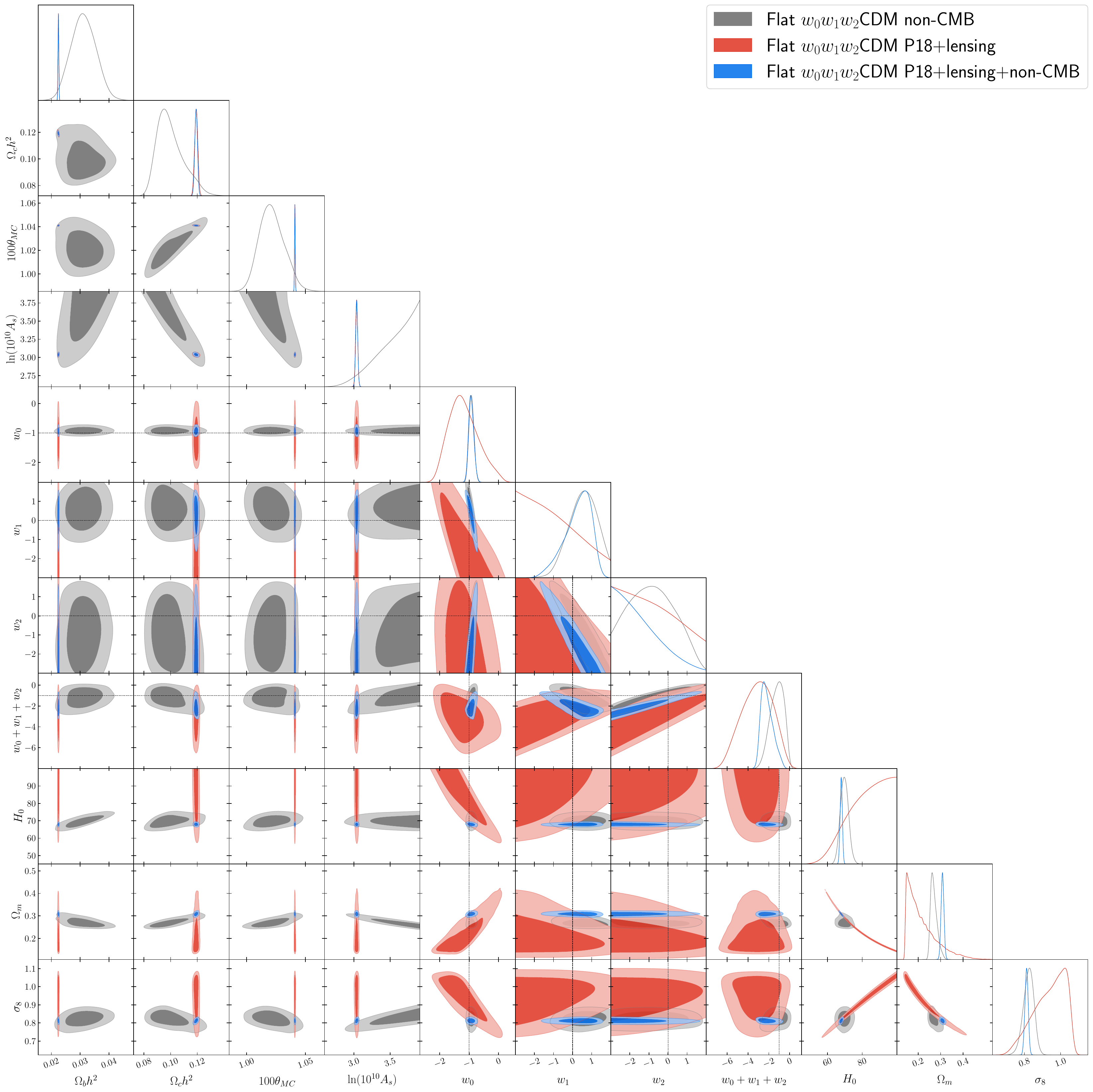}}
\caption{One-dimensional likelihoods and 1$\sigma$ and $2\sigma$ likelihood confidence contours of flat $w_0 w_1 w_2$CDM model parameters favored by non-CMB, P18+lensing, P18+lensing+non-CMB datasets. We do not show $\tau$ and $n_s$, which are fixed in the non-CMB data analysis.
}
\label{fig:flat_w0w1w2CDM_P18_lensing_nonCMB23v2}
\end{figure*}
\begin{figure*}[htbp]
\centering
\mbox{\includegraphics[width=127mm]{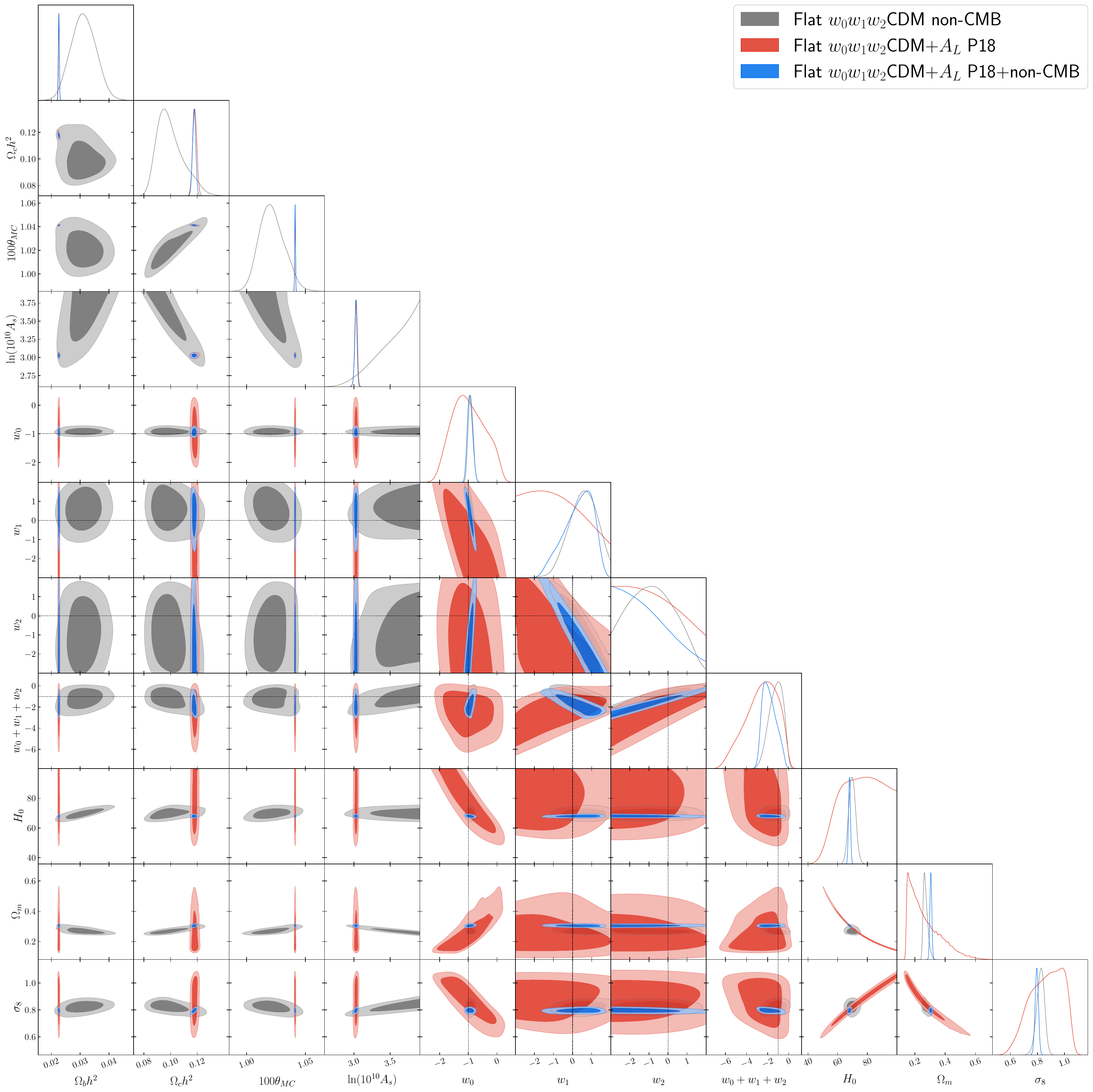}}
        \caption{One-dimensional likelihoods and 1$\sigma$ and $2\sigma$ likelihood confidence contours of flat $w_0 w_1 w_2$CDM$+A_L$ model parameters favored by non-CMB, P18, and P18+non-CMB datasets. We do not show $\tau$ and $n_s$, which are fixed in the non-CMB data analysis.
}
\label{fig:flat_w0w1w2CDM_Alens_P18_nonCMB23v2}
\end{figure*}
\begin{figure*}[htbp]
\centering
\mbox{\includegraphics[width=127mm]{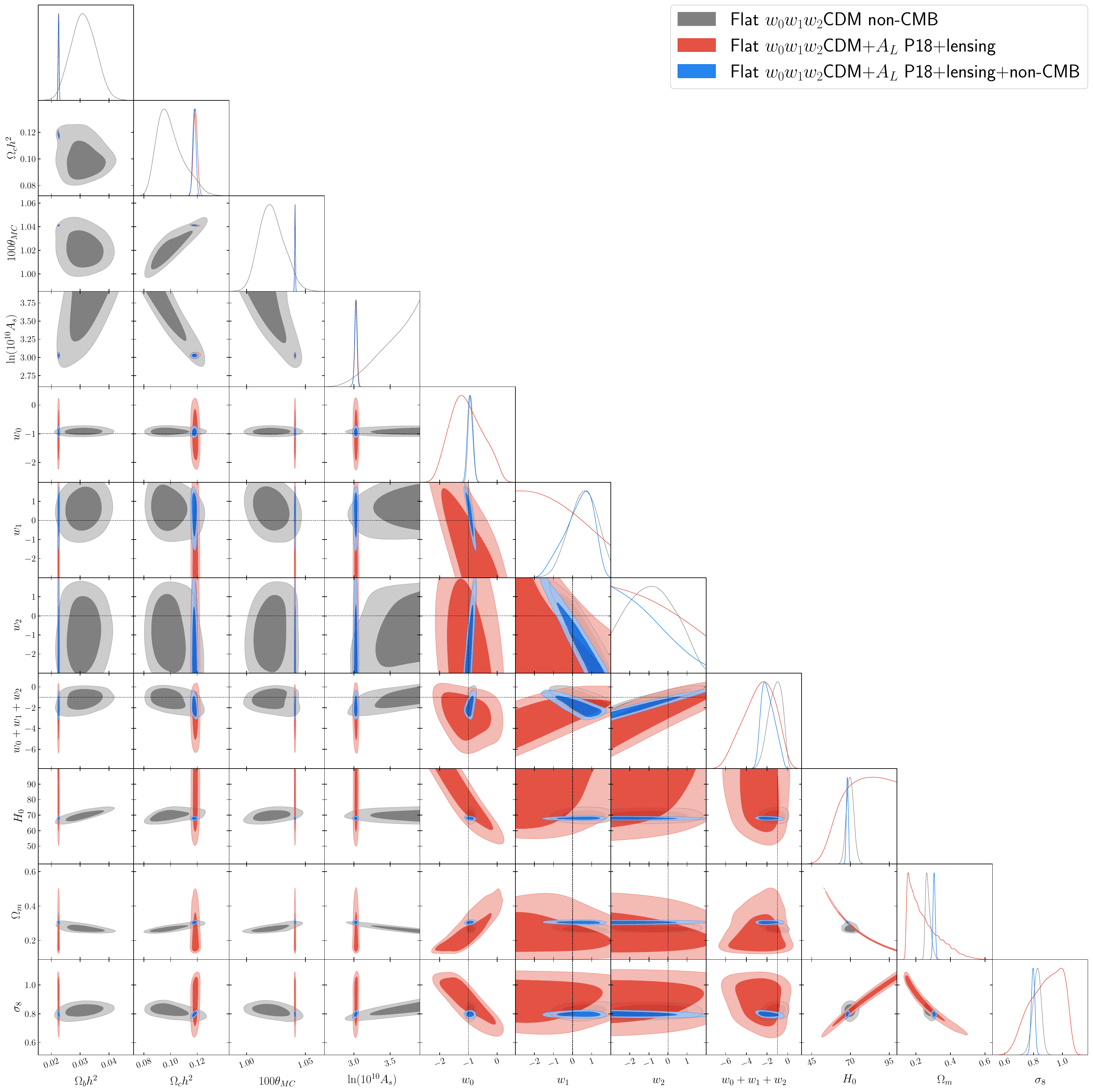}}
\caption{One-dimensional likelihoods and 1$\sigma$ and $2\sigma$ likelihood confidence contours of flat $w_0 w_1 w_2$CDM$+A_L$ model parameters favored by non-CMB, P18+lensing, P18+lensing+non-CMB datasets. We do not show $\tau$ and $n_s$, which are fixed in the non-CMB data analysis.
}
\label{fig:flat_w0w1w2CDM_Alens_P18_lensing_nonCMB23v2}
\end{figure*}
\begin{figure*}[htbp]
\centering
\mbox{\includegraphics[width=62mm]{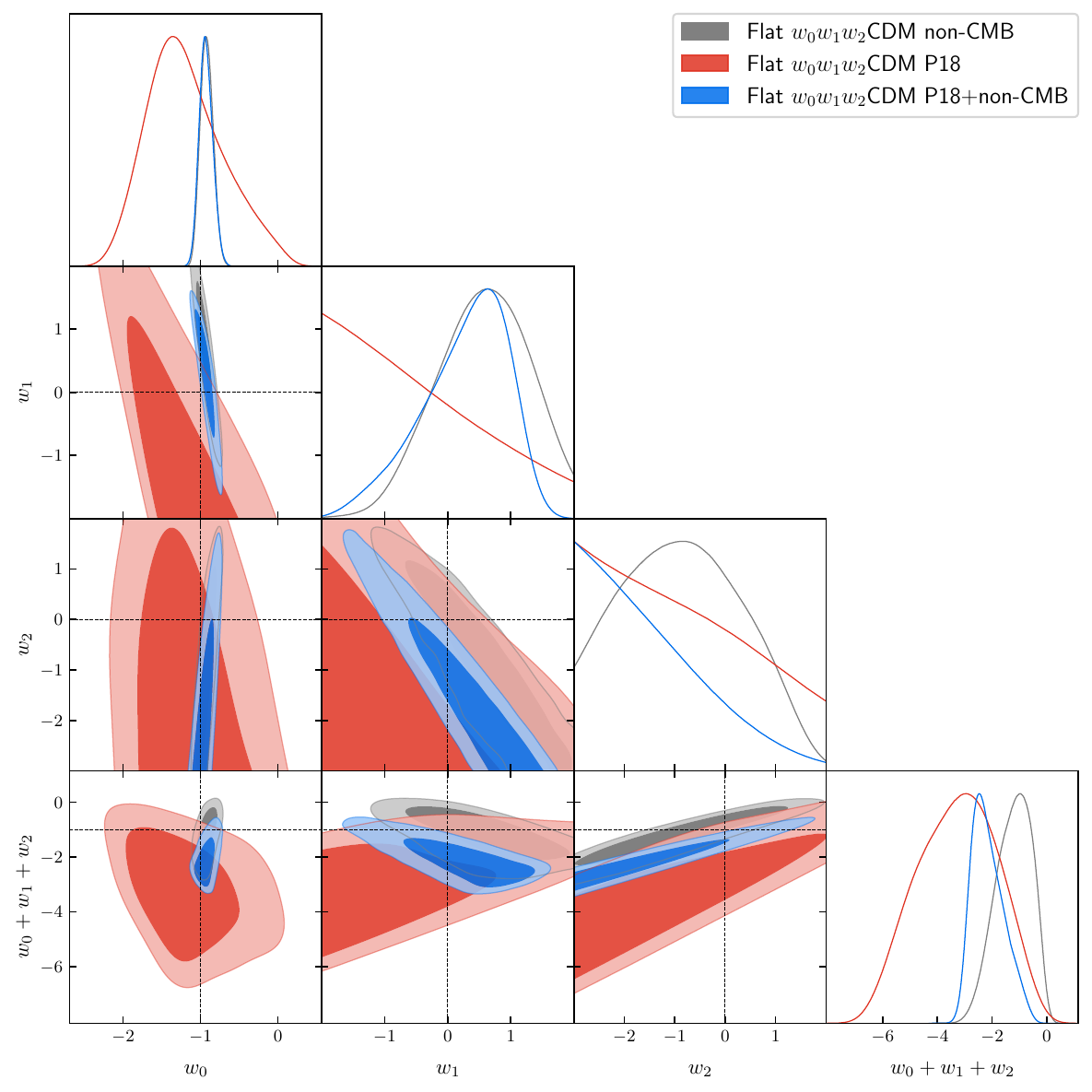}}
\mbox{\includegraphics[width=62mm]{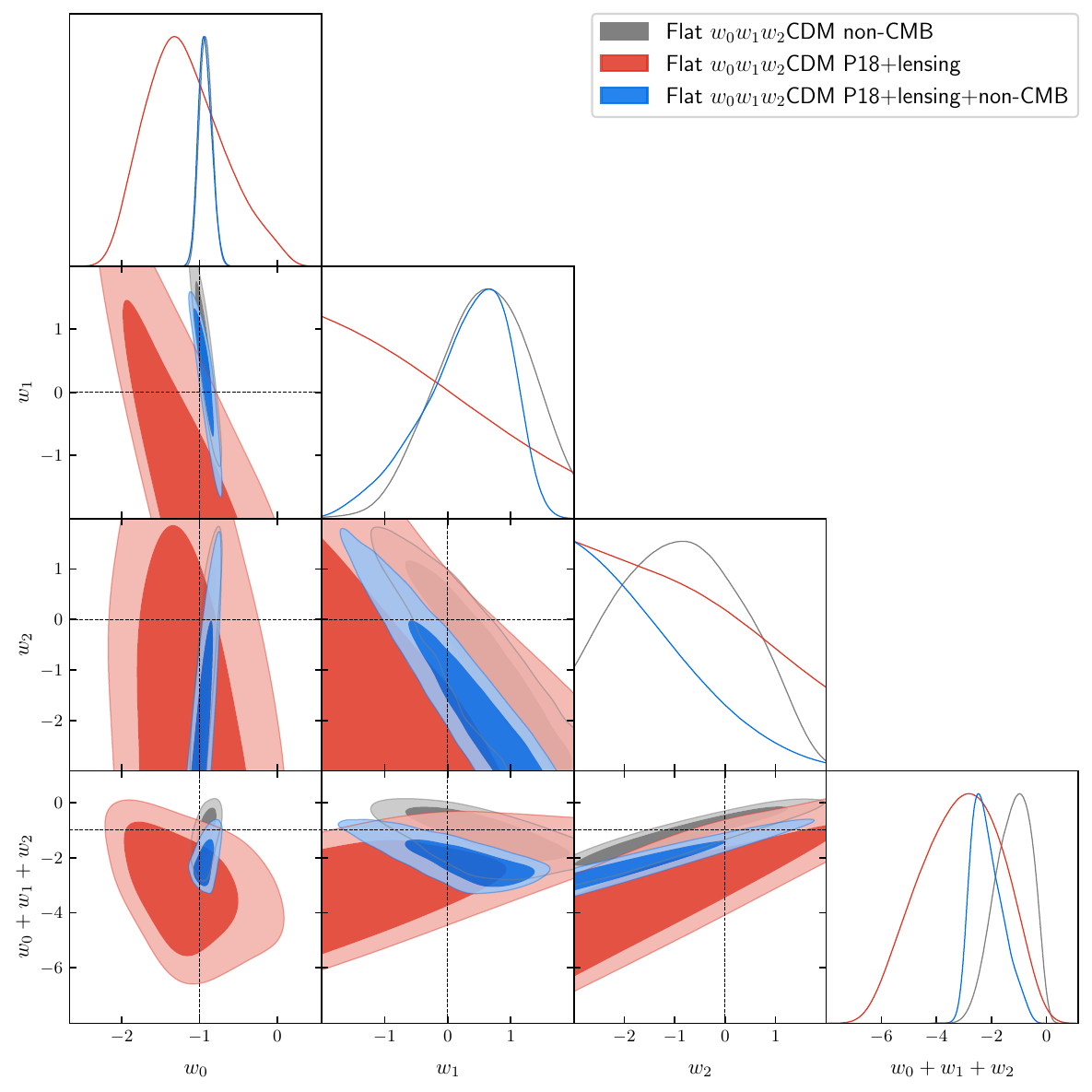}}
        \caption{One-dimensional likelihoods and 1$\sigma$ and $2\sigma$ likelihood confidence contours of $w_0$, $w_1$, $w_2$, and $w_0+w_1+w_2$ parameters in the flat $w_0 w_1 w_2$CDM parametrization favored by (left) non-CMB, P18, and P18+non-CMB datasets, and (right) non-CMB, P18+lensing, and P18+lensing+non-CMB datasets.
}
\label{fig:flat_w0w1w2CDM_P18_nonCMB23v2_w0w1w2}
\end{figure*}
\begin{figure*}[htbp]
\centering
\mbox{\includegraphics[width=62mm]{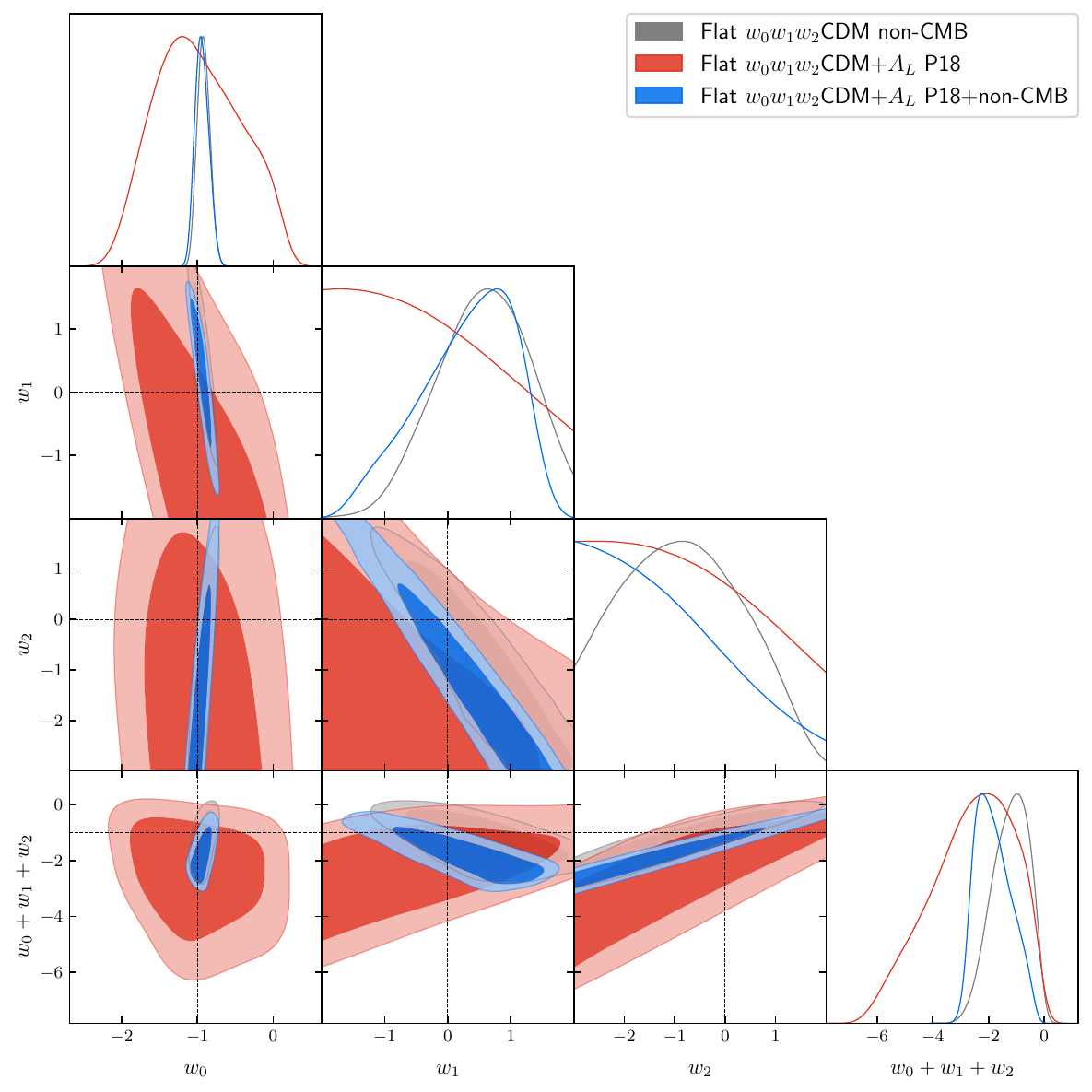}}
\mbox{\includegraphics[width=62mm]{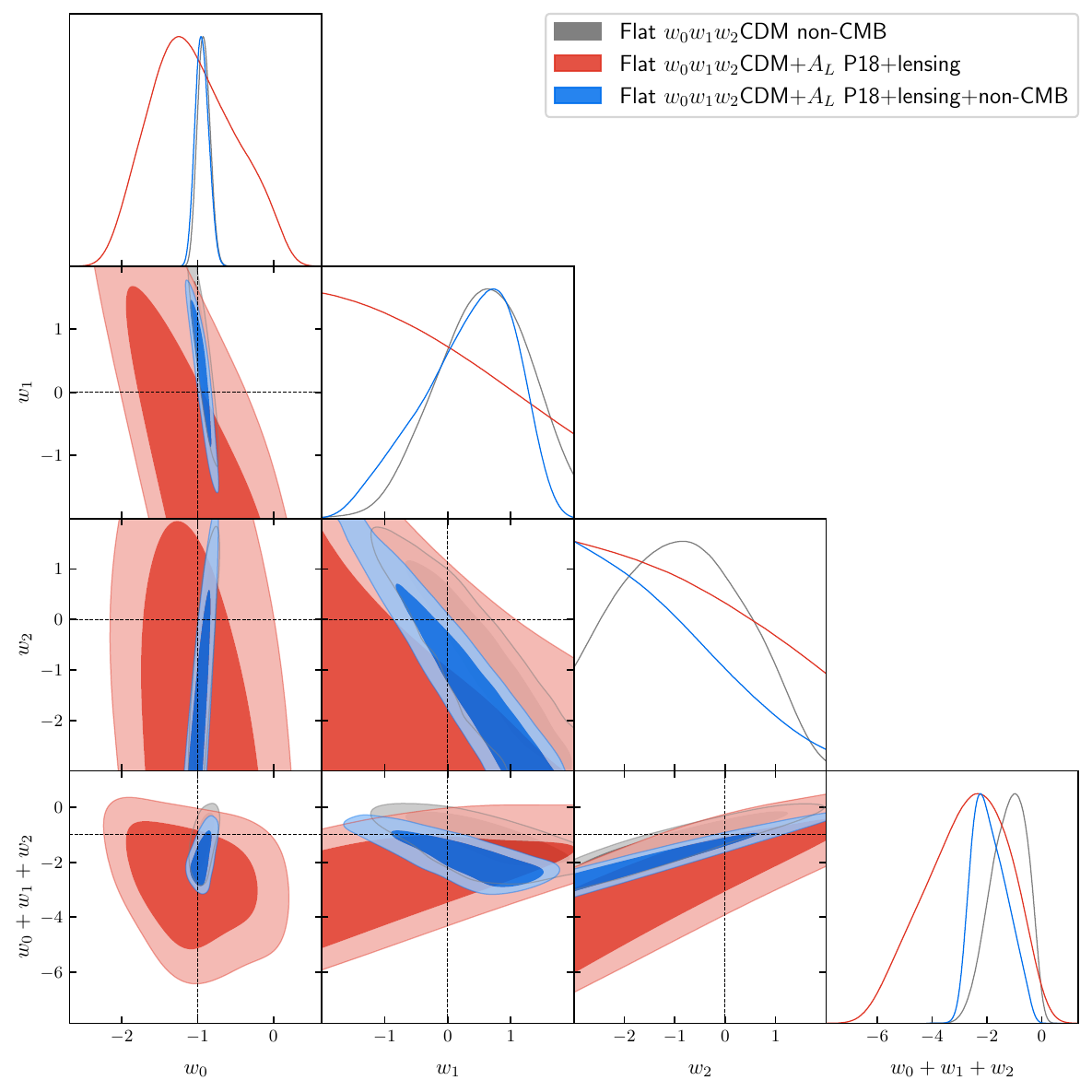}}
        \caption{One-dimensional likelihoods and 1$\sigma$ and $2\sigma$ likelihood confidence contours of $w_0$, $w_1$, $w_2$, and $w_0+w_1+w_2$ parameters in the flat $w_0 w_1 w_2$CDM$+A_L$ parametrization favored by (left) non-CMB, P18, and P18+non-CMB datasets, and (right) non-CMB, P18+lensing, and P18+lensing+non-CMB datasets.
}
\label{fig:flat_w0w1w2CDM_Alens_P18_nonCMB23v2_w0w1w2}
\end{figure*}

Tables \ref{tab:results_flat_w0w2CDM} and \ref{tab:results_flat_w0w2CDM_Alens} summarize the parameter constraints for the $w_0 w_2$CDM ($+A_L$) parametrizations. The likelihood distributions of the cosmological parameters are shown in Figures \ref{fig:flat_w0w2CDM_P18_nonCMB23v2}--\ref{fig:flat_w0w2CDM_Alens_P18_lensing_nonCMB23v2}, with just the $w_0$, $w_2$, and $w_0+w_2$ panels shown in Figures \ref {fig:flat_w0w2CDM_P18_nonCMB23v2_w0w2} and \ref{fig:flat_w0w2CDM_Alens_P18_nonCMB23v2_w0w2}. Tables \ref{tab:results_flat_w0w2p2CDM} and \ref{tab:results_flat_w0w2p2CDM_Alens} summarize the parameter constraints for the $w_0 w_p p$CDM ($+A_L$) parametrizations. The likelihood distributions of the cosmological parameters are shown in Figure \ref{fig:flat_w0w2p2CDM_P18_nonCMB23v2}--\ref{fig:flat_w0w2p2CDM_Alens_P18_lensing_nonCMB23v2}, with just the $w_0$, $w_p$, $p$, and $w_0+w_p$ panels shown in Figures \ref {fig:flat_w0w2p2CDM_P18_nonCMB23v2_w0w2p2} and \ref{fig:flat_w0w2p2CDM_Alens_P18_nonCMB23v2_w0w2p2}. Tables \ref{tab:results_flat_w0w1w2CDM} and \ref{tab:results_flat_w0w1w2CDM_Alens} summarize the parameter constraints for the $w_0 w_1 w_2$CDM ($+A_L$) parametrizations. The likelihood distributions of the cosmological parameters are shown in Figure \ref{fig:flat_w0w1w2CDM_P18_nonCMB23v2}--\ref{fig:flat_w0w1w2CDM_Alens_P18_lensing_nonCMB23v2}, with just the $w_0$, $w_1$, $w_2$, and $w_0+w_1+w_2$ panels shown in Figures \ref {fig:flat_w0w1w2CDM_P18_nonCMB23v2_w0w1w2} and \ref{fig:flat_w0w1w2CDM_Alens_P18_nonCMB23v2_w0w1w2}. Tables \ref{tab:results_flat_w0w2CDM}--\ref{tab:results_flat_w0w1w2CDM_Alens} also list the $\Delta\chi^2_{\rm min}$, $\Delta$AIC, and $\Delta$DIC values.

As in the $w_0w_a$CDM ($+A_L$) cases \cite{Chan-GyungPark:2024mlx, Chan-GyungPark:2024brx} the constraints on the $w(z)$ parameters, as well as on the derived parameters $H_0$, $\Omega_m$, and $\sigma_8$, from P18 and from P18+lensing data are less restrictive than those from non-CMB data.

We first compare the cosmological parameter values determined from P18+lensing+non-CMB data for the eight-parameter $w_0w_2$CDM and the nine-parameter $w_0w_2$CDM+$A_L$ parametrizations that are listed in the last columns of Tables \ref{tab:results_flat_w0w2CDM} and \ref{tab:results_flat_w0w2CDM_Alens}. Differences in the values of the six primary parameters common to the flat $\Lambda$CDM model are smaller than $1\sigma$: $\Omega_b h^2$ ($-0.54\sigma$), $\Omega_c h^2$ ($+0.82\sigma$), $100\theta_{\text{MC}}$ ($-0.37\sigma$), $\tau$ ($+0.42\sigma$), $n_s$ ($-0.67\sigma$), and $\ln(10^{10}A_s)$ ($+0.64\sigma$). For the two equation of state parameters, for the $w_0w_2$CDM parametrization we find $w_0=-0.898\pm 0.040$ and $w_2=-1.12^{+0.50}_{-0.40}$ while for the $w_0w_2$CDM+$A_L$ parametrization we have $w_0=-0.908\pm 0.040$ and $w_2=-0.78^{+0.48}_{-0.39}$, with the difference between these pair of values being $+0.18\sigma$ and $-0.54\sigma$. For the high-$z$ asymptotic value of $w(z)$, $w_0+w_2$, we find for the $w_0w_2$CDM parametrization $w_0+w_2=-2.02^{+0.47}_{-0.37}$ while for the $w_0w_2$CDM+$A_L$ parametrization $w_0+w_2=-1.69^{+0.45}_{-0.36}$, with a $-0.56\sigma$ difference between the two results. For the derived parameters $H_0$, $\Omega_m$, and $\sigma_8$, the differences are $+0.01\sigma$, $+0.28\sigma$, and $+1.03\sigma$, respectively.

   
\begin{table}[htbp]
\tbl{Mean and 68\% (or 95\%) confidence limits of flat $w_0 w_p p$CDM model parameters from non-CMB, P18, P18+lensing, P18+non-CMB, and P18+lensing+non-CMB data. $H_0$ has units of km s$^{-1}$ Mpc$^{-1}$. We also list the values of $\chi^2_{\text{min}}$, DIC, and AIC and the differences with respect to the values in the flat $\Lambda$CDM model for the same dataset, denoted by $\Delta\chi^2_{\text{min}}$, $\Delta$DIC, and $\Delta$AIC, respectively.}
{\begin{tabular}{@{}cccccc@{}} \toprule
  Parameter                     &  Non-CMB                     & P18                         &  P18+lensing             &  P18+non-CMB            & P18+lensing+non-CMB    \\
 \colrule
  $\Omega_b h^2$                & $0.0308 \pm 0.0044$          & $0.02239 \pm 0.00015$       & $0.02243 \pm 0.00015$    &  $0.02245 \pm 0.00014$  &  $0.02245 \pm 0.00014$ \\
  $\Omega_c h^2$                & $0.1014^{+0.0073}_{-0.012}$  & $0.1199 \pm 0.0014$         & $0.1193 \pm 0.0012$      &  $0.1191 \pm 0.0011$    &  $0.11904 \pm 0.00098$ \\
  $100\theta_\textrm{MC}$       & $1.0228^{+0.0097}_{-0.011}$  & $1.04094 \pm 0.00031$       & $1.04099 \pm 0.00031$    &  $1.04100 \pm 0.00030$  &  $1.04101 \pm 0.00030$ \\
  $\tau$                        & $0.0540$                     & $0.0540 \pm 0.0079$         & $0.0522 \pm 0.0075$      &  $0.0527 \pm 0.0077$    &  $0.0530 \pm 0.0073$  \\
  $n_s$                         & $0.9655$                     & $0.9655 \pm 0.0044$         & $0.9667 \pm 0.0041$      &  $0.9672 \pm 0.0039$    &  $0.9671 \pm 0.0038$  \\
  $\ln(10^{10} A_s)$            & $3.52\pm 0.26$ ($>3.06$)     & $3.044 \pm 0.016$           & $3.038 \pm 0.015$        &  $3.038 \pm 0.016$      &  $3.039 \pm 0.014$   \\
  $w_0$                         & $-0.865^{+0.048}_{-0.041}$   & $-1.40^{+0.29}_{-0.55}$     & $-1.39^{+0.26}_{-0.53}$  &  $-0.914^{+0.031}_{-0.046}$   &  $-0.916^{+0.031}_{-0.045}$ \\ 
  $w_p$                         & $-0.25^{+0.95}_{-0.23}$      & $-1.0 \pm 1.3$ ($< 1.36$)   & $-1.0 \pm 1.3$ ($<1.35$) &  $-1.70^{+0.52}_{-1.2}$       &  $-1.73^{+0.48}_{-1.2}$\\ 
  $p$                           & $2.5 \pm 1.1$ ($>0.454$)     & $1.93^{+0.87}_{-1.7}$ ($>0.257$) & $2.0 \pm 1.1$ ($>0.276$) &  $2.8 \pm 0.8 ($>1.20$)$ &  $2.86^{+1.1}_{-0.26}$ ($>1.31$)\\ 
 \colrule
  $w_0+w_p$                     & $-1.11^{+0.96}_{-0.19}$      & $-2.4^{+1.1}_{-1.3}$        & $-2.4^{+1.1}_{-1.3}$     &  $-2.61^{+0.54}_{-1.2}$ &  $-2.65^{+0.54}_{-1.1}$\\
  $H_0$                         & $69.7 \pm 2.4$               & $86 \pm 10$ ($> 66.8$)      & $85 \pm 10$ ($>66.8$)    &  $67.89 \pm 0.63$       &  $67.92 \pm 0.64$     \\
  $\Omega_m$                    & $0.273^{+0.011}_{-0.017}$    & $0.205^{+0.013}_{-0.062}$   & $0.205^{+0.014}_{-0.063}$&  $0.3086 \pm 0.0063$    &  $0.3082 \pm 0.0062$  \\
  $\sigma_8$                    & $0.818^{+0.031}_{-0.027}$    & $0.964^{+0.10}_{-0.045}$    & $0.955^{+0.10}_{-0.042}$ &  $0.811 \pm 0.011$      &  $0.8114 \pm 0.0091$  \\
 \colrule
  $\chi_{\textrm{min}}^2$       & $1457.48$                    & $2761.04$                   & $2770.34$                & $4231.80$               &  $4240.63$            \\
  $\Delta\chi_{\textrm{min}}^2$ & $-12.45$                     & $-4.76$                     & $-4.37$                  & $-8.44$                 &  $-8.63$              \\
  $\textrm{DIC}$                & $1471.29$                    & $2815.91$                   & $2824.51$                & $4289.09$               &  $4297.22$            \\
  $\Delta\textrm{DIC}$          & $-6.82$                      & $-2.02$                     & $-1.94$                  & $-3.24$                 &  $-3.98$             \\
  $\textrm{AIC}$                & $1471.48$                    & $2821.04$                   & $2830.34$                & $4291.80$               &  $4300.63$           \\
  $\Delta\textrm{AIC}$          & $-6.45$                      & $+1.24$                     & $+1.63$                  & $-2.44$                 &  $-2.63$            \\
\botrule
\end{tabular}\label{tab:results_flat_w0w2p2CDM}}
\end{table}


   
\begin{table}[htbp]
\tbl{Mean and 68\% (or 95\%) confidence limits of flat $w_0 w_p p$CDM$+A_L$ model parameters from non-CMB, P18, P18+lensing, P18+non-CMB, and P18+lensing+non-CMB data. $H_0$ has units of km s$^{-1}$ Mpc$^{-1}$. We also list the values of $\chi^2_{\text{min}}$, DIC, and AIC and the differences with respect to the values in the flat $\Lambda$CDM model for the same dataset, denoted by $\Delta\chi^2_{\text{min}}$, $\Delta$DIC, and $\Delta$AIC, respectively.}
{\begin{tabular}{@{}cccccc@{}} \toprule
  Parameter                     &  Non-CMB                     & P18                         &  P18+lensing                 &  P18+non-CMB                 & P18+lensing+non-CMB    \\
 \colrule
  $\Omega_b h^2$                & $0.0308 \pm 0.0044$          & $0.02258 \pm 0.00017$       & $0.02250 \pm 0.00017$        &  $0.02263 \pm 0.00015$       &  $0.02255 \pm 0.00015$ \\
  $\Omega_c h^2$                & $0.1014^{+0.0073}_{-0.012}$  & $0.1182 \pm 0.0015$         & $0.1185 \pm 0.0015$          &  $0.1177 \pm 0.0012$         &  $0.1178 \pm 0.0012$ \\
  $100\theta_\textrm{MC}$       & $1.0228^{+0.0097}_{-0.011}$  & $1.04113 \pm 0.00032$       & $1.04108 \pm 0.00032$        &  $1.04118 \pm 0.00031$       &  $1.04113 \pm 0.00030$ \\
  $\tau$                        & $0.0540$                     & $0.0495^{+0.0089}_{-0.0077}$& $0.0494^{+0.0087}_{-0.0074}$ &  $0.0481^{+0.0088}_{-0.0074}$&  $0.0481^{+0.0087}_{-0.0073}$ \\
  $n_s$                         & $0.9655$                     & $0.9705 \pm 0.0049$         & $0.9690 \pm 0.0048$          &  $0.9718 \pm 0.0043$         &  $0.9705 \pm 0.0042$  \\
  $\ln(10^{10} A_s)$            & $3.52\pm 0.26$ ($>3.06$)     & $3.030^{+0.019}_{-0.016}$   & $3.029^{+0.019}_{-0.016}$    &  $3.026^{+0.018}_{-0.015}$   &  $3.025^{+0.018}_{-0.015}$  \\
  $A_L$                         & \ldots                       & $1.165^{+0.063}_{-0.088}$   & $1.047^{+0.039}_{-0.058}$    &  $1.184 \pm 0.064$           &  $1.072 \pm 0.038$   \\
  $w_0$                         & $-0.865^{+0.048}_{-0.041}$   & $-1.16^{+0.44}_{-0.66}$     & $-1.24^{+0.37}_{-0.63}$      &  $-0.916^{+0.032}_{-0.043}$  &  $-0.917^{+0.032}_{-0.042}$ \\
  $w_p$                         & $-0.25^{+0.95}_{-0.23}$      & $-0.81^{+0.82}_{-2.1}$      & $-0.8 \pm 1.3$ ($<1.41$)     &  $-1.29^{+1.1}_{-0.69}$      &  $-1.34^{+1.1}_{-0.69}$\\ 
  $p$                           & $2.5 \pm 1.1$ ($>0.454$)     & $2.0^{+1.3}_{-1.5}$         & $2.0 \pm 1.1$ ($>0.269$)     &  $2.8 \pm 0.9 ($>0.980$)$    &  $2.8 \pm 0.9$ ($>1.01$)\\ 
 \colrule
  $w_0+w_p$                     & $-1.11^{+0.96}_{-0.19}$      & $-1.97^{+1.5}_{-0.95}$      & $-2.0^{+1.4}_{-1.1}$         &  $-2.20^{+1.1}_{-0.62}$      &  $-2.26^{+1.2}_{-0.63}$\\
  $H_0$                         & $69.7 \pm 2.4$               & $77^{+20}_{-8}$ ($> 54.7$)  & $80 \pm 13$ ($>57.7$)        &  $67.94 \pm 0.65$            &  $67.91 \pm 0.64$     \\
  $\Omega_m$                    & $0.273^{+0.011}_{-0.017}$    & $0.261^{+0.030}_{-0.012}$   & $0.241^{+0.023}_{-0.10}$     &  $0.3054 \pm 0.0063$         &  $0.3059 \pm 0.0063$  \\
  $\sigma_8$                    & $0.818^{+0.031}_{-0.027}$    & $0.874^{+0.16}_{-0.080}$    & $0.899^{+0.15}_{-0.067}$     &  $0.795 \pm 0.012$           &  $0.797 \pm 0.012$  \\
 \colrule
  $\chi_{\textrm{min}}^2$       & $1457.48$                    & $2755.89$                   & $2770.20$                    & $4221.77$                    &  $4237.05$            \\
  $\Delta\chi_{\textrm{min}}^2$ & $-12.45$                     & $-9.91$                     & $-4.51$                      & $-18.47$                     &  $-12.21$              \\
  $\textrm{DIC}$                & $1471.29$                    & $2812.89$                   & $2825.96$                    & $4282.60$                    &  $4295.60$            \\
  $\Delta\textrm{DIC}$          & $-6.82$                      & $-5.04$                     & $-0.49$                      & $-9.73$                      &  $-5.60$             \\
  $\textrm{AIC}$                & $1471.48$                    & $2817.89$                   & $2832.20$                    & $4282.60$                    &  $4299.05$           \\
  $\Delta\textrm{AIC}$          & $-6.45$                      & $-1.91$                     & $+3.50$                      & $-10.47$                     &  $-4.21$            \\
\botrule
\end{tabular}\label{tab:results_flat_w0w2p2CDM_Alens}}
\end{table}


We now compare the cosmological parameter values determined from P18+lensing+non-CMB data for the nine-parameter $w_0w_pp$CDM and the ten-parameter $w_0w_pp$CDM+$A_L$ parametrizations that are listed in the last columns of Tables \ref{tab:results_flat_w0w2p2CDM} and \ref{tab:results_flat_w0w2p2CDM_Alens}. Differences in the values of the six primary parameters common to the flat $\Lambda$CDM model are smaller than $1\sigma$: $\Omega_b h^2$ ($-0.49\sigma$), $\Omega_c h^2$ ($+0.80\sigma$), $100\theta_{\text{MC}}$ ($-0.28\sigma$), $\tau$ ($+0.43\sigma$), $n_s$ ($-0.60\sigma$), and $\ln(10^{10}A_s)$ ($+0.61\sigma$). For the three equation of state parameters, for the $w_0w_pp$CDM parametrization we find $w_0=-0.916^{+0.031}_{-0.045}$, $w_p=-1.73^{+0.48}_{-1.2}$, and $p = 2.86^{+1.1}_{-0.26}\ (> 1.31)$ while for the $w_0w_pp$CDM+$A_L$ parametrization we have $w_0=-0.917^{+0.032}_{-0.042}$, $w_p=-1.34^{+1.1}_{-0.69}$, and $p = 2.8 \pm 0.9\ (> 1.01)$, with the difference between these triplets of values being $+0.02\sigma$, $-0.46\sigma$, and $+0.06\sigma$. For the high-$z$ asymptotic value of $w(z)$, $w_0+w_p$, we find for the $w_0w_pp$CDM parametrization $w_0+w_p=-2.65^{+0.54}_{-1.1}$ while for the $w_0w_pp$CDM+$A_L$ parametrization $w_0+w_p=-2.26^{+1.2}_{-0.63}$, with a $-0.47\sigma$ difference between the two results. For the derived parameters $H_0$, $\Omega_m$, and $\sigma_8$, the differences are $+0.01\sigma$, $+0.26\sigma$, and $+0.96\sigma$, respectively. 

We finally compare the cosmological parameter values determined from P18+lensing+non-CMB data for the nine-parameter $w_0w_1w_2$CDM and the ten-parameter $w_0w_1w_2$CDM+$A_L$ parametrizations that are listed in the last columns of Tables \ref{tab:results_flat_w0w1w2CDM} and \ref{tab:results_flat_w0w1w2CDM_Alens}. Differences in the values of the six primary parameters common to the flat $\Lambda$CDM model are smaller than $1\sigma$: $\Omega_b h^2$ ($-0.49\sigma$), $\Omega_c h^2$ ($+0.85\sigma$), $100\theta_{\text{MC}}$ ($-0.33\sigma$), $\tau$ ($+0.42\sigma$), $n_s$ ($-0.64\sigma$), and $\ln(10^{10}A_s)$ ($+0.64\sigma$). For the three equation of state parameters, for the $w_0w_1w_2$CDM parametrization we find $w_0=-0.929^{+0.077}_{-0.095}$, $w_1=0.27^{+0.87}_{-0.45}$, and $w_2 = -1.5 \pm 1.1\ (< 0.709)$ while for the $w_0w_1w_2$CDM+$A_L$ parametrization we have $w_0=-0.943^{+0.083}_{-0.095}$, $w_1=0.31^{+0.94}_{-0.54}$, and $w_2 = -1.2 \pm 1.2\ (< 1.08)$, with the difference between these triplets of values being $+0.11\sigma$, $-0.04\sigma$, and $-0.18\sigma$. For the high-$z$ asymptotic value of $w(z)$, $w_0+w_1+w_2$, we find for the $w_0w_1w_2$CDM parametrization $w_0+w_1+w_2=-2.15^{+0.40}_{-0.71}$ while for the $w_0w_1w_2$CDM+$A_L$ parametrization $w_0+w_1+w_2=-1.88^{+0.50}_{-0.79}$, with a $-0.30\sigma$ difference between the two results. For the derived parameters $H_0$, $\Omega_m$, and $\sigma_8$, the differences are $-0.04\sigma$, $+0.34\sigma$, and $+0.99\sigma$, respectively.

   
\begin{table}[htbp]
\tbl{Mean and 68\% (or 95\%) confidence limits of flat $w_0 w_1 w_2$CDM model parameters
        from non-CMB, P18, P18+lensing, P18+non-CMB, and P18+lensing+non-CMB data.
        $H_0$ has units of km s$^{-1}$ Mpc$^{-1}$. We also list the values of $\chi^2_{\text{min}}$, DIC, and AIC and the differences with respect to the values in the flat $\Lambda$CDM model for the same dataset, denoted by $\Delta\chi^2_{\text{min}}$, $\Delta$DIC, and $\Delta$AIC, respectively.}
{\begin{tabular}{@{}cccccc@{}} \toprule
  Parameter                     &  Non-CMB                     & P18                         &  P18+lensing             &  P18+non-CMB            & P18+lensing+non-CMB    \\
 \colrule
  $\Omega_b h^2$                & $0.0311 \pm 0.0043$          & $0.02240 \pm 0.00015$       & $0.02244 \pm 0.00015$    &  $0.02244 \pm 0.00014$  &  $0.02245 \pm 0.00014$ \\
  $\Omega_c h^2$                & $0.0999^{+0.0064}_{-0.012}$  & $0.1198 \pm 0.0014$         & $0.1192 \pm 0.0012$      &  $0.1191 \pm 0.0011$    &  $0.11912 \pm 0.00099$ \\
  $100\theta_\textrm{MC}$       & $1.0214^{+0.0092}_{-0.012}$  & $1.04094 \pm 0.00031$       & $1.04101 \pm 0.00031$    &  $1.04100 \pm 0.00030$  &  $1.04099 \pm 0.00030$ \\
  $\tau$                        & $0.0540$                     & $0.0540 \pm 0.0078$         & $0.0520 \pm 0.0075$      &  $0.0524 \pm 0.0077$    &  $0.0528 \pm 0.0074$  \\
  $n_s$                         & $0.9656$                     & $0.9656 \pm 0.0044$         & $0.9671 \pm 0.0042$      &  $0.9671 \pm 0.0040$    &  $0.9670 \pm 0.0038$  \\
  $\ln(10^{10} A_s)$            & $3.56\pm 0.26$ ($>3.06$)     & $3.043 \pm 0.016$           & $3.037 \pm 0.015$        &  $3.038 \pm 0.016$      &  $3.039 \pm 0.014$   \\
  $w_0$                         & $-0.920^{+0.078}_{-0.088}$   & $-1.19^{+0.39}_{-0.59}$     & $-1.18^{+0.40}_{-0.59}$  &  $-0.928^{+0.078}_{-0.094}$& $-0.929^{+0.077}_{-0.095}$  \\  
  $w_1$                         & $0.54^{+0.83}_{-0.66}$       & $-1.1 \pm 1.3$ ($< 1.29$)   & $-1.1 \pm 1.3$ ($<1.36$) &  $0.25^{+0.86}_{-0.47}$    & $0.27^{+0.87}_{-0.45}$\\ 
  $w_2$                         & $-0.8^{+1.1}_{-1.4}$         & $-0.9 \pm 1.4$ ($< 1.51$)   & $-1.1 \pm 1.3$ ($<1.54$) &  $-1.5 \pm 1.1$ ($<0.732$) & $-1.5 \pm 1.1$ ($<0.709$)\\ 
 \colrule
  $w_0+w_1+w_2$                 & $-1.21^{+0.78}_{-0.56}$      & $-3.3^{+1.6}_{-1.5}$        & $-3.1^{+1.7}_{-1.4}$     &  $-2.14^{+0.42}_{-0.73}$&  $-2.15^{+0.40}_{-0.71}$\\
  $H_0$                         & $69.8 \pm 2.3$               & $84 \pm 11$ ($> 63.4$)      & $83 \pm 11$ ($>63.6$)    &  $67.92 \pm 0.64$       &  $67.92 \pm 0.64$     \\
  $\Omega_m$                    & $0.2703^{+0.0093}_{-0.018}$  & $0.217^{+0.018}_{-0.074}$   & $0.218^{+0.019}_{-0.075}$&  $0.3084 \pm 0.0063$    &  $0.3084 \pm 0.0063$  \\
  $\sigma_8$                    & $0.823^{+0.031}_{-0.026}$    & $0.950^{+0.12}_{-0.053}$    & $0.939^{+0.11}_{-0.052}$ &  $0.812 \pm 0.011$      &  $0.8118 \pm 0.0090$  \\
 \colrule
  $\chi_{\textrm{min}}^2$       & $1457.13$                    & $2760.82$                   & $2770.51$                & $4232.22$               &  $4240.75$            \\
  $\Delta\chi_{\textrm{min}}^2$ & $-12.80$                     & $-4.98$                     & $-4.20$                  & $-8.02$                 &  $-8.51$              \\
  $\textrm{DIC}$                & $1471.28$                    & $2815.65$                   & $2824.37$                & $4289.73$               &  $4298.65$            \\
  $\Delta\textrm{DIC}$          & $-6.83$                      & $-2.28$                     & $-2.08$                  & $-2.60$                 &  $-2.55$             \\
  $\textrm{AIC}$                & $1471.13$                    & $2820.82$                   & $2830.51$                & $4292.22$               &  $4300.75$           \\
  $\Delta\textrm{AIC}$          & $-6.83$                      & $+1.02$                     & $+1.80$                  & $-2.02$                 &  $-2.51$            \\
\botrule
\end{tabular}\label{tab:results_flat_w0w1w2CDM}}
\end{table}

   
\begin{table}[htbp]
\tbl{Mean and 68\% (or 95\%) confidence limits of flat $w_0 w_1 w_2$CDM$+A_L$ model parameters
        from non-CMB, P18, P18+lensing, P18+non-CMB, and P18+lensing+non-CMB data.
        $H_0$ has units of km s$^{-1}$ Mpc$^{-1}$. We also list the values of $\chi^2_{\text{min}}$, DIC, and AIC and the differences with respect to the values in the flat $\Lambda$CDM model for the same dataset, denoted by $\Delta\chi^2_{\text{min}}$, $\Delta$DIC, and $\Delta$AIC, respectively.}
{\begin{tabular}{@{}cccccc@{}} \toprule
  Parameter                     &  Non-CMB                     & P18                         &  P18+lensing             &  P18+non-CMB            & P18+lensing+non-CMB    \\
 \colrule
  $\Omega_b h^2$                & $0.0311 \pm 0.0043$          & $0.02259 \pm 0.00017$       & $0.02249 \pm 0.00017$    &  $0.02263 \pm 0.00016$  &  $0.02255 \pm 0.00015$ \\
  $\Omega_c h^2$                & $0.0999^{+0.0064}_{-0.012}$  & $0.1181 \pm 0.0015$         & $0.1185 \pm 0.0015$      &  $0.1176 \pm 0.0012$    &  $0.1178 \pm 0.0012$ \\
  $100\theta_\textrm{MC}$       & $1.0214^{+0.0092}_{-0.012}$  & $1.04115 \pm 0.00033$       & $1.04107 \pm 0.00033$    &  $1.04117 \pm 0.00031$  &  $1.04113 \pm 0.00030$ \\
  $\tau$                        & $0.0540$                     & $0.0494^{+0.0087}_{-0.0075}$& $0.0491^{+0.0086}_{-0.0074}$ &  $0.0482^{+0.0085}_{-0.0072}$ & $0.0481^{+0.0085}_{-0.0074}$  \\
  $n_s$                         & $0.9656$                     & $0.9706 \pm 0.0049$         & $0.9688 \pm 0.0048$      &  $0.9718 \pm 0.0043$       &  $0.9706 \pm 0.0042$  \\
  $\ln(10^{10} A_s)$            & $3.56\pm 0.26$ ($>3.06$)     & $3.030^{+0.018}_{-0.016}$   & $3.029^{+0.018}_{-0.016}$&  $3.026^{+0.017}_{-0.015}$ &  $3.025^{+0.017}_{-0.015}$   \\
  $A_L$                         & \ldots                       & $1.158^{+0.060}_{-0.087}$   & $1.040^{+0.041}_{-0.053}$&  $1.185 \pm 0.064$         &  $1.073^{+0.036}_{-0.040}$   \\
  $w_0$                         & $-0.920^{+0.078}_{-0.088}$   & $-0.99^{+0.52}_{-0.67}$     & $-1.06^{+0.50}_{-0.66}$  &  $-0.940^{+0.082}_{-0.10}$ & $-0.943^{+0.083}_{-0.095}$  \\  
  $w_1$                         & $0.54^{+0.83}_{-0.66}$       & $-0.8 \pm 1.3$ ($< 1.51$)   & $-0.9 \pm 1.4$ ($<1.51$) &  $0.29^{+0.97}_{-0.54}$    & $0.31^{+0.94}_{-0.54}$\\ 
  $w_2$                         & $-0.8^{+1.1}_{-1.4}$         & $-0.8 \pm 1.4$ ($< 1.55$)   & $-0.8 \pm 1.4$ ($<1.56$) &  $-1.2 \pm 1.2$ ($<1.18$)  & $-1.2 \pm 1.2$ ($<1.08$)\\ 
 \colrule
  $w_0+w_1+w_2$                 & $-1.21^{+0.78}_{-0.56}$      & $-2.6^{+1.9}_{-1.1}$        & $-2.7^{+1.8}_{-1.3}$     &  $-1.82^{+0.47}_{-0.82}$   &  $-1.88^{+0.50}_{-0.79}$\\
  $H_0$                         & $69.8 \pm 2.3$               & $77^{+20}_{-10}$ ($> 54.9$) & $79 \pm 13$ ($>57.6$)    &  $67.96 \pm 0.65$          &  $67.96 \pm 0.64$     \\
  $\Omega_m$                    & $0.2703^{+0.0093}_{-0.018}$  & $0.267^{+0.036}_{-0.13}$    & $0.249^{+0.030}_{-0.11}$ &  $0.3051 \pm 0.0064$       &  $0.3054 \pm 0.0063$  \\
  $\sigma_8$                    & $0.823^{+0.031}_{-0.026}$    & $0.868^{+0.16}_{-0.088}$    & $0.891^{+0.14}_{-0.078}$ &  $0.795 \pm 0.012$         &  $0.797 \pm 0.012$  \\
 \colrule
  $\chi_{\textrm{min}}^2$       & $1457.13$                    & $2755.91$                   & $2770.25$                & $4222.96$               &  $4237.21$            \\
  $\Delta\chi_{\textrm{min}}^2$ & $-12.80$                     & $-9.89$                     & $-4.46$                  & $-17.28$                &  $-12.05$        \\
  $\textrm{DIC}$                & $1471.28$                    & $2813.02$                   & $2832.25$                & $4282.32$               &  $4296.50$        \\
  $\Delta\textrm{DIC}$          & $-6.83$                      & $-4.91$                     & $-0.17$                  & $-10.01$                &  $-4.70$         \\
  $\textrm{AIC}$                & $1471.13$                    & $2817.91$                   & $2832.25$                & $4284.96$               &  $4299.21$       \\
  $\Delta\textrm{AIC}$          & $-6.83$                      & $-1.89$                     & $+3.54$                  & $-9.28$                 &  $-4.05$         \\
\botrule
\end{tabular}\label{tab:results_flat_w0w1w2CDM_Alens}}
\end{table}

To assess how relatively-well these spatially-flat dynamical dark energy parametrizations do in fitting the P18+lensing+non-CMB dataset, we compare their DIC values, more precisely their $\Delta$DIC values relative to the DIC value of the standard flat $\Lambda$CDM model. These $\Delta$DIC values are listed in Tables \ref{tab:results_flat_w0w2CDM}--\ref{tab:results_flat_w0w1w2CDM_Alens} here, with the flat $w_0 w_a$CDM ($+A_L$) values listed in tables 1 and 2 of Ref.\ \refcite{Chan-GyungPark:2024brx}, the flat XCDM$+A_L$ parametrization value listed in table XI of Ref.\ \refcite{deCruzPerez:2024abc} (in the flat XCDM parametrization with $A_L = 1$ the P18+lensing and the non-CMB cosmological constraints are inconsistent at $3.6\sigma$, ruling the parametrization out at this significance, see Ref.\ \refcite{deCruzPerez:2024abc}), and the flat $\Lambda$CDM$+A_L$ value listed in table VII of Ref.\ \refcite{deCruzPerez:2024abc}. There is {\it positive} evidence in favor of all models and parametrizations relative to the flat $\Lambda$CDM model, with the $w_0 w_p p$CDM$+A_L$ parametrization with $\Delta {\rm DIC} = -5.60$ favored the most and the $w_0 w_a$CDM parametrization with $\Delta {\rm DIC} = -2.45$ favored the least over the flat $\Lambda$CDM model.    

We now investigate whether the largest data compilation we study, the P18+lensing+non-CMB dataset, provides model-independent cosmological parameter constraints by comparing these constraints in several dark energy models and parametrizations. The P18+lensing+non-CMB data parameter constraints for the flat $\Lambda$CDM ($+A_L$) models are in tables IV and VII of Ref.\ \refcite{deCruzPerez:2024abc}, while those for the flat XCDM ($+A_L$) parametrization are in table XI there. In this analysis we do not consider the flat XCDM parametrization with $A_L=1$ since P18+lensing constraints and non-CMB constraints are more than $3\sigma$ inconsistent in this case \cite{deCruzPerez:2024abc}. P18+lensing+non-CMB data parameter constraints for the flat $w_0 w_a$CDM ($+A_L$) parametrizations are in table 1 and 2  of Ref.\ \refcite{Chan-GyungPark:2024brx}, while those of the flat $w_0 w_2$CDM ($+A_L$), $w_0 w_p p$CDM ($+A_L$), and $w_0 w_1 w_2$CDM ($+A_L$) parametrizations are in Tables \ref{tab:results_flat_w0w2CDM} --\ref{tab:results_flat_w0w1w2CDM_Alens}.

We first consider the $A_L = 1$ flat dark energy models and parametrizations and first compute the largest model-to-model differences for the six common model parameters. For $\Omega_b h^2$ the largest value is $0.02249 \pm 0.00013$ in the $\Lambda$CDM model and the smallest is $0.02244 \pm 0.00014$ in the $w_0 w_a$CDM and $w_0 w_2$CDM parametrizations, resulting in a difference of $0.26\sigma$. For $\Omega_c h^2$ the smallest value is $0.11849 \pm 0.00084$ in the $\Lambda$CDM model and the largest is $0.11917 \pm 0.00099$ in the $w_0 w_2$CDM parametrization, resulting in a difference of $0.52\sigma$. For $100\theta_\textrm{MC}$ the $\Lambda$CDM model $100\theta_\textrm{MC}=1.04109 \pm 0.00028$ is the largest and the $w_0 w_2$CDM parametrization $100\theta_\textrm{MC}=1.04098 \pm 0.00030$ is the smallest, giving a difference of $0.27\sigma$. For $\tau$ and $n_s$ the largest deviation is $0.41\sigma$ and $0.32\sigma$, respectively, between the largest $\Lambda$CDM model values ($\tau=0.0569 \pm 0.0071$ and $n_s=0.9685 \pm 0.0036$) and the smallest $w_0 w_2$CDM parametrization values ($\tau=0.0527 \pm 0.0074$ and $n_s=0.9668 \pm 0.0038$). For $\ln (10^{10} A_s)$ the $\Lambda$CDM model largest value ($\ln (10^{10} A_s)=3.046 \pm 0.014$) and the $w_0 w_2$CDM, $w_0 w_p p$CDM, and $w_0 w_1 w_2$CDM parametrizations smallest value ($\ln (10^{10} A_s)=3.039 \pm 0.014$) differ at $0.35\sigma$. Thus for $A_L = 1$ the six common model parameters agree between the various dark energy models and parametrizations to well within $1\sigma$, and we expect the same for the derived parameters, as we now show. For the derived parameter $H_0$, $68.05 \pm 0.38$ km s$^{-1}$ Mpc$^{-1}$ in the $\Lambda$CDM model is the largest value and $67.80 \pm 0.64$ km s$^{-1}$ Mpc$^{-1}$ in the $w_0 w_a$CDM parametrization is the smallest, giving a difference of $0.34\sigma$. The $\Lambda$CDM model has the smallest $\Omega_m=0.3059 \pm 0.0050$ and the $w_0 w_a$CDM parametrization has the largest $\Omega_m=0.3094 \pm 0.0063$ which differ from each other by $0.44\sigma$ while the $\Lambda$CDM model smallest $\sigma_8=0.8077 \pm 0.0057$ and the $w_0 w_2$CDM parameterization largest $\sigma_8=0.8125 \pm 0.0091$ differ by $0.45\sigma$. In the $A_L = 1$ case the P18+lensing+non-CMB data compilation provides reasonably model-independent parameter constraints, with the largest difference being the $0.52\sigma$ for $\Omega_c h^2$.

For the $A_L$-varying dark energy models and parametrizations, where we include the XCDM$+A_L$ case, the largest model-to-model differences for the six common model parameters are as follows. For $\Omega_b h^2$ the largest is $0.02263 \pm 0.00014$ in the XCDM+$A_L$ parametrization and the smallest is $0.02255 \pm 0.00015$ in the $w_0 w_2$CDM+$A_L$ or $w_0 w_p p$CDM+$A_L$ or $w_0 w_1 w_2$CDM+$A_L$ parametrization, resulting in a difference of $0.39\sigma$. For $\Omega_c h^2$ the smallest is $0.1168 \pm 0.0011$ in the XCDM$+A_L$ parametrization and the largest is $0.1179 \pm 0.0012$ in the $w_0 w_2$CDM$+A_L$ case, resulting in a difference of $0.68\sigma$. For $100\theta_\textrm{MC}$, the XCDM$+A_L$ parametrization $100\theta_\textrm{MC}=1.04126 \pm 0.00030$ is the largest and the $w_0 w_p p$CDM$+A_L$ or $w_0 w_1 w_2$CDM$+A_L$ case $100\theta_\textrm{MC}=1.04113 \pm 0.00030$ is the smallest, giving a difference of $0.31\sigma$. For $\tau$ the largest difference is $0.16\sigma$ between the $\Lambda$CDM+$A_L$ model smallest $\tau=0.0477 \pm 0.0086$ and the XCDM+$A_L$ parametrization largest $\tau=0.0496 \pm 0.0083$, while for $n_s$ the largest difference is $0.48\sigma$ between the XCDM+$A_L$ parametrization largest $n_s=0.9733 \pm 0.0040$ and the $w_0 w_p p$CDM+$A_L$ case smallest $n_s=0.9705 \pm 0.0042$. For $\ln (10^{10} A_s)$, the $\Lambda$CDM+$A_L$ model smallest $\ln (10^{10} A_s)=3.024 \pm 0.018$ and the XCDM+$A_L$ case largest $\ln (10^{10} A_s)=3.026 \pm 0.017$ differ at $0.08\sigma$. For the derived parameters, the largest $H_0=68.45 \pm 0.42$ km s$^{-1}$ Mpc$^{-1}$ in the $\Lambda$CDM+$A_L$ model and the smallest $H_0=67.79 \pm 0.63$ km s$^{-1}$ Mpc$^{-1}$ in the XCDM+$A_L$ case differ at $0.87\sigma$. The $\Lambda$CDM+$A_L$ model smallest $\Omega_m=0.3005 \pm 0.0053$ and the $w_0 w_a$CDM+$A_L$ case largest $\Omega_m=0.3062 \pm 0.0064$ differ by $0.69\sigma$ while the XCDM+$A_L$ model smallest $\sigma_8=0.785 \pm 0.011$ and the $w_0 w_2$CDM+$A_L$ or $w_0 w_p p$CDM or $w_0 w_1 w_2$CDM parametrization largest $\sigma_8=0.797 \pm 0.012$ differ from each other by $0.74\sigma$. The spread in parameter values across dark energy models and parametrizations, when measured using P18+lensing+non-CMB data, is a little larger when $A_L$ is allowed to vary, compared to the $A_L = 1$ case. Among the six common primary parameters, the biggest spread, $0.68\sigma$, is again in $\Omega_ch^2$, while the three derived parameters have larger spreads, ranging from $0.69\sigma$ to $0.87\sigma$. Since these are all smaller than $1\sigma$ we conclude that even in the varying $A_L$ case P18+lensing+non-CMB data provide reasonably model-independent cosmological parameter measurements.

For all the $A_L=1$ and $A_L$-varying dark energy models and parametrizations together, but excluding the XCDM case with $A_L=1$, the largest model-to-model differences for the six common model parameters are as follows. For $\Omega_b h^2$, the largest $0.02263 \pm 0.00014$ in the XCDM+$A_L$ case and the smallest $0.02244 \pm 0.00014$ in the $w_0 w_a$CDM or $w_0 w_2$CDM parametrization have a difference of $0.96\sigma$. For $\Omega_c h^2$ the smallest is $0.1168 \pm 0.0011$ in the XCDM$+A_L$ case and the largest is $0.11917 \pm 0.00099$ in the $w_0 w_2$CDM parametrization, resulting in a difference of $1.60\sigma$. For $100\theta_\textrm{MC}$, the XCDM+$A_L$ case $100\theta_\textrm{MC}=1.04126 \pm 0.00030$ is the largest and the $w_0 w_2$CDM parametrization $100\theta_\textrm{MC}=1.04098 \pm 0.00030$ is the smallest, and these have a difference of $0.66\sigma$. For $\tau$ the largest difference is $0.82\sigma$ between the $\Lambda$CDM model largest $\tau=0.0569 \pm 0.0071$ and the $\Lambda$CDM+$A_L$ model smallest $\tau=0.0477 \pm 0.0086$, while for $n_s$ the largest difference is $1.18\sigma$ between the XCDM+$A_L$ case largest $n_s=0.9733 \pm 0.0040$ and the $w_0 w_2$CDM case smallest $n_s=0.9668 \pm 0.0038$. For $\ln (10^{10} A_s)$, the $\Lambda$CDM model largest $\ln (10^{10} A_s)=3.046 \pm 0.014$ and the $\Lambda$CDM+$A_L$ model smallest $\ln (10^{10} A_s)=3.024 \pm 0.018$ differ at $0.96\sigma$. For the derived parameters, the largest $H_0=68.45 \pm 0.42$ km s$^{-1}$ Mpc$^{-1}$ in the $\Lambda$CDM+$A_L$ model and the smallest $H_0=67.79 \pm 0.63$ km s$^{-1}$ Mpc$^{-1}$ in the XCDM+$A_L$ case differ by $0.87\sigma$. The $\Lambda$CDM+$A_L$ model smallest $\Omega_m=0.3005 \pm 0.0053$ and the $w_0 w_a$CDM case largest $\Omega_m=0.3094 \pm 0.0063$ have the largest difference of $1.08\sigma$ while the XCDM+$A_L$ case smallest $\sigma_8=0.785 \pm 0.011$ and the $w_0 w_2$CDM model parametrization largest $\sigma_8=0.8125 \pm 0.0091$ differ by $1.93\sigma$. The parameter value differences are larger when we compare across all the $A_L = 1$ and varying $A_L$ cases, caused by the somewhat significant changes in some parameter values when the lensing consistency parameter is allowed to vary, compared to the $A_L = 1$ case values.  

As described above, the significance of the difference between the largest and smallest $H_0$ values are $0.34\sigma$ (when $A_L = 1$), $0.87\sigma$ (when $A_L$ varies), and $0.87\sigma$ (overall), indicating that for these models P18+lensing+non-CMB data provide a reasonably model-independent $H_0$ value, which may be summarized as $H_0 = 68.1 \pm 0.9$ km~s$^{-1}$~Mpc$^{-1}$, where the central value is the average of the smallest and largest values and the error bar is half of the spread between the largest upper $1\sigma$ value and the smallest lower $1\sigma$ value. This summary values agrees with the median statistics result $H_0=68\pm 2.8$ km s$^{-1}$ Mpc$^{-1}$ \cite{Chen:2011ab, Gottetal2001, Calabreseetal2012}. It also agrees with some local measurements including the summary value of Ref.\ \refcite{Cao:2023eja} $H_0=69.25\pm 2.42$ km s$^{-1}$ Mpc$^{-1}$ from a joint analysis of $H(z)$, BAO, Pantheon+ SNIa, quasar angular size, reverberation-measured \mii\ and \civ\ quasar, and 118 Amati correlation gamma-ray burst data, and the local $H_0=69.03\pm 1.75$ km s$^{-1}$ Mpc$^{-1}$ from JWST TRGB+JAGB and SNIa data \cite{Freedman:2024eph}, but is smaller than the local $H_0=73.04\pm 1.04$ km s$^{-1}$ Mpc$^{-1}$ measured using Cepheids and SNIa data \cite{Riess:2021jrx}, also see Refs.\ \citen{Chen:2024gnu, Barua:2024gei}.

Table \ref{tab:results_quphAL} lists the current low-redshift, $w_0$, and high-redshift, $w(z \rightarrow \infty)$, values of the dynamical dark energy equation of state parameter, $w(z)$, as well as the lensing consistency parameter, $A_L$, values, measured using P18+lensing+non-CMB data, in a number of dynamical dark energy parametrizations. The XCDM$+A_L$ parametrization value is from Ref.\ \refcite{deCruzPerez:2024abc} and the $w_0 w_a$CDM ($+A_L$) values are from Refs.\ \citen{Chan-GyungPark:2024mlx, Chan-GyungPark:2024brx}.


\begin{table}[htbp]
\tbl{Mean and 68\% confidence limits of $w_0$, $w(z \rightarrow \infty)$, and $A_L$ values in dynamical dark energy parametrizations for P18+lensing+non-CMB data. Significances shown in parentheses for $w_0$ and $w(z \rightarrow \infty)$ indicate how much larger (smaller) than $-1$ they are when the following index is qu (ph), while those for $A_L$ indicate how much larger than unity it is. We do not list values for the XCDM parameterization as that is observationally inconsistent at $>3\sigma$ significance with these data.}
{\begin{tabular}{@{}lccc@{}} \toprule
  Parametrization        & $w_0$                                          &  $w(z \rightarrow \infty)$                 & $A_L$     \\
 \colrule
  XCDM$+A_L$             & $-0.968 \pm 0.024$ ($1.33\sigma$ qu)           & $-0.968 \pm 0.024$ ($1.33\sigma$ qu)       & $1.101 \pm 0.037$ ($2.73\sigma$)   \\
  $w_0 w_a$CDM           & $-0.850 \pm 0.059$ ($2.54\sigma$ qu)           &  $-1.44^{+0.20}_{-0.17}$ ($2.20\sigma$ ph) & 1   \\            
  $w_0 w_a$CDM$+A_L$     & $-0.879 \pm 0.060$ ($2.02\sigma$ qu)           &  $-1.27^{+0.20}_{-0.17}$ ($1.35\sigma$ ph) & $1.078^{+0.036}_{-0.040}$ ($1.95\sigma$)  \\              
  $w_0 w_2$CDM           & $-0.898 \pm 0.040$ ($2.55\sigma$ qu)           &  $-2.02^{+0.47}_{-0.37}$ ($2.17\sigma$ ph) & 1   \\                 
  $w_0 w_2$CDM$+A_L$     & $-0.908 \pm 0.040$ ($2.30\sigma$ qu)           &  $-1.69^{+0.45}_{-0.36}$ ($1.53\sigma$ ph) & $1.072 \pm 0.038$ ($1.89\sigma$)  \\           
  $w_0 w_p p$CDM         & $-0.916^{+0.031}_{-0.045}$ ($1.87\sigma$ qu)   &  $-1.73^{+0.48}_{-1.2}$ ($1.52\sigma$ ph)  & 1   \\        
  $w_0 w_p p$CDM$+A_L$   & $-0.917^{+0.032}_{-0.042} $ ($1.98\sigma$ qu)  &  $-2.26^{+1.2}_{-0.63}$ ($1.05\sigma$ ph)  & $1.072 \pm 0.038$ ($1.89\sigma$)  \\        
  $w_0 w_1 w_2$CDM       & $-0.929^{+0.077}_{-0.095}$ ($0.747\sigma$ qu)  &  $-2.15^{+0.40}_{-0.71}$ ($2.88\sigma$ ph) & 1   \\            
  $w_0 w_1 w_2$CDM$+A_L$ & $-0.943^{+0.083}_{-0.095} $ ($0.600\sigma$ qu) &  $-1.88^{+0.50}_{-0.79}$ ($1.76\sigma$ ph) & $1.073^{+0.036}_{-0.040}$ ($1.83\sigma$)  \\         
\botrule
\end{tabular}\label{tab:results_quphAL}}
\end{table}


In the XCDM parametrization P18+lensing data favor phantom-like ($w< -1$) dynamical dark energy while non-CMB data favor quintessence-like ($w> -1$) behavior \cite{deCruzPerez:2024abc}, so much so that the cosmological constraints from these two datasets are inconsistent at $3.6\sigma$, thus ruling out the XCDM parametrization at $> 3\sigma$ \cite{deCruzPerez:2024abc}. Allowing $A_L$ to be an additional free parameter to be determined from data, so now considering the XCDM$+A_L$ dynamical dark energy parametrization, somewhat reconciles the P18+lensing and the non-CMB constraints, to $2.4\sigma$ inconsistency, and a joint P18+lensing+non-CMB data analysis then indicates a $1.3\sigma$ preference for a quintessence-like $w_0 = -0.968 \pm 0.024$, but also requires $A_L = 1.101 \pm 0.037$, $>1$ at $2.73\sigma$ significance \cite{deCruzPerez:2024abc}. As noted above, non-CMB data are more effective at constraining $w(z)$ than are P18 or P18+lensing data. In the XCDM parametrization, non-CMB data results in $w_0 = -0.853^{+0.043}_{-0.033}$, while P18+lensing data give $w_0 = -1.55 \pm 0.26$ ($w_0 = -1.34^{+0.26}_{-0.51}$) when $A_L = 1$ (when $A_L$ is allowed to vary and determined from P18+lensing data to be $1.054 \pm 0.055$). This shift in $w_0$ towards quintessence-like behavior when $A_L$ is allowed to vary away from unity suggests that the observed excess smoothing of some of the Planck CMB multipoles (relative to what is expected in the best-fit Planck cosmological model) is partially responsible for the phantom-like behavior seen in the XCDM parametrization when P18+lensing data are used in the analysis.

When $w(z)$ is parametrized using more than one parameter, such as $w_0$ and $w_a$ in the $w_0 w_a$CDM parametrization where $w(z) = w_0 + w_a z/(1+z)$, there is more freedom so the situation differs a bit from the XCDM case discussed in the previous paragraph. In the two and three parameter $w(z)$ cases we study the P18+lensing constraints and the non-CMB constraints are inconsistent at less than $3\sigma$, even when $A_L = 1$, see Table \ref{tab:consistency_w(z)CDM} and the related discussion above; this differs from what happens in the XCDM case. However, allowing $A_L$ to vary makes the P18+lensing constraints and the non-CMB constraints more consistent, which is similar to what happens in the XCDM case. In the $w_0 w_a$CDM case, as discussed in Ref.\ \refcite{Chan-GyungPark:2024brx}, non-CMB data favor quintessence-like behavior, with $w_0 = -0.876 \pm 0.055$, while P18+lensing data favor phantom-like behavior, with $w_0 = -1.24^{+0.44}_{-0.56}$ when $A_L = 1$ and $w_0 = -1.14^{+0.48}_{-0.68}$ with $A_L = 1.046^{+0.038}_{-0.057}$. In this case P18+lensing+non-CMB data favors quintessence-like $w_0 = -0.850 \pm 0.059$ at $2.54\sigma$ ($w_0 = -0.879 \pm 0.060$ at $2.02\sigma$) while favoring quintessence-like $w(z \rightarrow \infty) = w_0 + w_a = -1.44^{+0.20}_{-0.17}$ at $2.20\sigma$ when $A_L= 1$ ($w(z \rightarrow \infty) = -1.27^{+0.20}_{-0.17}$ at $1.35\sigma$ with $A_L= 1.078^{+0.036}_{-0.040}$, $> 1$ at $1.95\sigma$), see discussion in Ref.\ \refcite{Chan-GyungPark:2024brx}. 

Similar behavior is true for all the other dynamical dark energy parametrizations we study here, as can be seen from the numerical values listed in Table \ref{tab:results_quphAL}. Excluding XCDM$+A_L$, from the results for the other four parametrizations listed in Table \ref{tab:results_quphAL}, when $A_L = 1$ $w(z \rightarrow \infty)$ is phantom-like at significance ranging from a low of $1.5\sigma$ to a high of $2.9\sigma$ (with three of the four parametrizations favoring phantom-like behavior at $> 2\sigma$), while when $A_L$ is allowed to vary the support for phantom-like $w(z \rightarrow \infty)$ has reduced significance ranging from a low of $1.1\sigma$ to a high of $1.8\sigma$ (with none of the four parametrizations favoring phantom-like behavior at $> 2\sigma$).  

We noted above that the two- and three-parameter $w(z)$ parametrizations have more flexibility than the XCDM parametrization and that at low and high redshift the two- and three-parameter parametrizations behave like XCDM parametrizations, but with two different constant $w$ parameters. We also noted above that non-CMB data is more effective at constraining $w(z)$ since dark energy is not as important at the higher redshift, $z \sim 1100$, where CMB data is more sensitive. In the XCDM parametrization this means that in a joint analysis, like P18+lensing+non-CMB, which is only possible when $A_L$ is allowed to vary (and which brings the P18+lensing constraints and the non-CMB constraints into less than $3\sigma$ inconsistency), non-CMB data overwhelms P18+lensing data and results in quintessence-like dynamical dark energy, even though P18+lensing data favor phantom-like dynamical dark energy. In the two- and three-parameter $w(z)$CDM parametrizations, $w(z)$ has more flexibility and while non-CMB data is able to ensure low-redshift $w_0$ values that are quintessence-like even in joint P18+lensing+non-CMB analyses, and also increase the P18+lensing data high-redshift $w(z \rightarrow \infty)$ values, they are unable to pull it into the more-physical quintessence-like regime, although in all cases they are able to reduce the evidence for phantom-like behavior to below $2\sigma$.

In addition to what is established from the numerical results given in Table \ref{tab:consistency_w(z)CDM}, we can see by comparing the corresponding panels in Figures \ref{fig:flat_w0w2CDM_P18_nonCMB23v2_w0w2} and \ref{fig:flat_w0w2CDM_Alens_P18_nonCMB23v2_w0w2}, \ref{fig:flat_w0w2p2CDM_P18_nonCMB23v2_w0w2p2} and \ref{fig:flat_w0w2p2CDM_Alens_P18_nonCMB23v2_w0w2p2}, and \ref{fig:flat_w0w1w2CDM_P18_nonCMB23v2_w0w1w2} and \ref{fig:flat_w0w1w2CDM_Alens_P18_nonCMB23v2_w0w1w2}, that allowing $A_L$ to vary makes the P18 or the P18+lensing cosmological constraints more consistent with the non-CMB ones, compared to the $A_L = 1$ case. In what follows we focus on the P18+lensing data (red contours) and the non-CMB data (gray contours), those shown in the right-hand panels of these figures. In the $w_0-w_2$ subpanels of the right-hand panels of Figures \ref{fig:flat_w0w2CDM_P18_nonCMB23v2_w0w2} and \ref{fig:flat_w0w2CDM_Alens_P18_nonCMB23v2_w0w2}, for the $w_0w_2$CDM ($+A_L$) parametrizations, one sees when $A_L=1$
(Fig.~\ref{fig:flat_w0w2CDM_P18_nonCMB23v2_w0w2}) the $2\sigma$ red and gray contours have some overlap, but when $A_L$ is allowed to vary (Fig.~\ref{fig:flat_w0w2CDM_Alens_P18_nonCMB23v2_w0w2}) the $2\sigma$ grey contours has some overlap with the $1\sigma$ red contour and vice versa, with the gray and red $1\sigma$ contours almost touching. Similarly, in the $w_0 - w_p$, $w_0 - p$, and $w_p - p$ right-hand subpanels of Figures \ref{fig:flat_w0w2p2CDM_P18_nonCMB23v2_w0w2p2} and \ref{fig:flat_w0w2p2CDM_Alens_P18_nonCMB23v2_w0w2p2} for the $w_0w_pp$CDM ($+A_L$) parametrizations we see a significantly improved consistency between the red and gray contours when going from the $A_L = 1$ case to the varying $A_L$ case; there is a degeneracy in the $w_p - p$ subpanels constraint contours that makes it difficult to examine this issue there. This improved consistency is also seen in the $w_0 - w_1$, $w_0 - w_2$, and $w_1 - w_2$ right-hand subpanels of Figures \ref{fig:flat_w0w1w2CDM_P18_nonCMB23v2_w0w1w2} and \ref{fig:flat_w0w1w2CDM_Alens_P18_nonCMB23v2_w0w1w2} for the $w_0w_1w_2$CDM ($+A_L$) parametrizations, most prominently in the $w_0 - w_1$ subpanels. 

From the numerical results for the $A_L = 1$ cases in Table \ref{tab:results_quphAL}, we see that at high-$z$ the dynamical dark energy equation of state parameter $w(z \rightarrow \infty)$ is phantom-like at a significance of $1.52\sigma$ ($w_0w_pp$CDM), $2.17\sigma$ ($w_0w_2$CDM), $2.20\sigma$ ($w_0w_a$CDM), and $2.88\sigma$ ($w_0w_1w_2$CDM) for P18+lensing+non-CMB data. This support for dynamical dark energy over a cosmological constant can also be seen in Figures \ref{fig:flat_w0w2CDM_P18_nonCMB23v2_w0w2}, \ref{fig:flat_w0w2p2CDM_P18_nonCMB23v2_w0w2p2}, and \ref{fig:flat_w0w1w2CDM_P18_nonCMB23v2_w0w1w2}; for the $w_0w_a$CDM parametrization see Ref.\ \refcite{Chan-GyungPark:2024mlx}. We focus here on P18+lensing+non-CMB data, and so on the blue contours in the right-hand panels in these figures. In the $w_0-w_2$ right-hand subpanel in Figure \ref{fig:flat_w0w2CDM_P18_nonCMB23v2_w0w2} for the $w_0w_2$CDM parametrization we see that the $\Lambda$CDM model point at $w_0 = -1$ and $w_2 = 0$ lies outside the $2\sigma$ blue contour. The same is true for the $w_0 - w_p$, $w_0 - p$, and $w_p - p$ right-hand subpanels of Figure \ref{fig:flat_w0w2p2CDM_P18_nonCMB23v2_w0w2p2} for the $w_0w_pp$CDM parametrization, where the $\Lambda$CDM model corresponds to the point $w_0 = -1$ and $w_p = 0$, $w_0 = -1$ and $p = 0$, and $w_p = 0$ and $p = 0$, respectively, with the $\Lambda$CDM model points lying outside the $2\sigma$ blue contours in all three subpanels. In the right-hand $w_0 - w_1$, $w_0 - w_2$, and $w_1 - w_2$ subpanels of Figure \ref{fig:flat_w0w1w2CDM_P18_nonCMB23v2_w0w1w2} for the $w_0w_1w_2$CDM parametrization, the $\Lambda$CDM model point $w_0 = -1$, $w_1 = 0$, and $w_2 = 0$ touches the $2\sigma$ blue contour in the $w_0 - w_1$ subpanel while lying outside the $2\sigma$ blue contours in the  $w_0 - w_2$ and $w_1 - w_2$ subpanels.

From the numerical results for the varying $A_L$ cases in Table \ref{tab:results_quphAL}, we see that at high-$z$ the dynamical dark energy equation of state parameter $w(z \rightarrow \infty)$ is phantom-like at a significance of $1.05\sigma$ ($w_0w_pp$CDM$+A_L$), $1.35\sigma$ ($w_0w_a$CDM$+A_L$), $1.53\sigma$ ($w_0w_2$CDM$+A_L$), and $1.76\sigma$ ($w_0w_1w_2$CDM$+A_L$) for P18+lensing+non-CMB data, with $A>1$ at a significance ranging from $1.83\sigma$ to $1.95\sigma$ depending on parametrization. This reduced support for dynamical dark energy over a cosmological constant (compared to the corresponding $A_L = 1$ case) can also be seen in Figures \ref{fig:flat_w0w2CDM_Alens_P18_nonCMB23v2_w0w2}, \ref{fig:flat_w0w2p2CDM_Alens_P18_nonCMB23v2_w0w2p2}, and \ref{fig:flat_w0w1w2CDM_Alens_P18_nonCMB23v2_w0w1w2}; for the $w_0w_a$CDM$+A_L$ parametrization see Ref.\ \refcite{Chan-GyungPark:2024brx}. We again focus on P18+lensing+non-CMB data and so on the blue contours in the right-hand panels in these figures. In the $w_0-w_2$ right-hand subpanel in Figure \ref{fig:flat_w0w2CDM_Alens_P18_nonCMB23v2_w0w2} for the $w_0w_2$CDM$+A_L$ parametrization we see that the $\Lambda$CDM model point at $w_0 = -1$ and $w_2 = 0$ lies between the $1\sigma$ and $2\sigma$ blue contours. The same is true for the $w_0 - w_p$, $w_0 - p$, and $w_p - p$ right-hand subpanels of Figure \ref{fig:flat_w0w2p2CDM_Alens_P18_nonCMB23v2_w0w2p2} for the $w_0w_pp$CDM$+A_L$ parametrization, where the $\Lambda$CDM model corresponds to the point $w_0 = -1$ and $w_p = 0$, $w_0 = -1$ and $p = 0$, and $w_p = 0$ and $p = 0$, respectively, with the $\Lambda$CDM model points lying between the $1\sigma$ and $2\sigma$ blue contours, but closer to the $1\sigma$ ones, in all three subpanels. In the right-hand $w_0 - w_1$, $w_0 - w_2$, and $w_1 - w_2$ subpanels of Figure \ref{fig:flat_w0w1w2CDM_Alens_P18_nonCMB23v2_w0w1w2} for the $w_0w_1w_2$CDM$+A_L$ parametrization, the $\Lambda$CDM model point $w_0 = -1$, $w_1 = 0$, and $w_2 = 0$ touches the $2\sigma$ blue contours in the $w_0 - w_1$ and $w_0 - w_2$ subpanels while lying between the $1\sigma$ and $2\sigma$ blue contours in the $w_1 - w_2$ subpanel.

In summary, unlike in the flat $\Lambda$CDM ($+A_L$) models, tables IV and VII in Ref.\ \refcite{deCruzPerez:2024abc}, in the dynamical dark energy parametrizations we consider, including the flat XCDM$+A_L$ parametrization, table XI of Ref.\ \refcite{deCruzPerez:2024abc}, and the flat $w_0w_a$CDM ($+A_L$) parametrizations \cite{Chan-GyungPark:2024mlx, Chan-GyungPark:2024brx}, and the flat $w_0w_2$CDM ($+A_L$), $w_0w_pp$CDM ($+A_L$), and $w_0w_1w_2$CDM ($+A_L$) parametrizations, Tables \ref{tab:results_flat_w0w2CDM}--\ref{tab:results_flat_w0w1w2CDM_Alens}, high-$z$ P18 and P18+lensing data favor phantom-like dynamical dark energy, higher $H_0$ values, and smaller $\Omega_m$ values, all with larger error bars, while low-$z$ non-CMB data favor quintessence-like dynamical dark energy, lower $H_0$ values, and larger $\Omega_m$ values, all with smaller error bars. Joint analysis of P18+lensing+non-CMB data breaks the parameter degeneracy and compromises by picking a slightly lower $H_0$ value and a slightly larger $\Omega_m$ value than are favored by non-CMB data but with much more restrictive error bars. Excluding the XCDM$+A_L$ case, in the two- and three-parameter $w(z)$CDM cases the joint P18+lensing+non-CMB analysis results in a less, but still, phantom-like $w(z \rightarrow \infty) < -1$ when $A_L = 1$, which becomes even less, but still, phantom-like when $A_L$ is allowed to vary and be simultaneously determined from these data. This evidence for phantom-like dark energy in these parametrizations is neither due to DESI BAO data (which we do not use), nor is it due to Pantheon+ or other SNIa data (see Refs.\ \citen{Chan-GyungPark:2024mlx, Chan-GyungPark:2024brx} for analysis and discussion in the $w_0w_a$CDM case). Because all the parametrizations we study reduce to the XCDM parametrization at high $z$, all the parametrizations favor phantom-like dynamical dark energy at larger $z$. When $A_L = 1$ they favor phantom-like behavior at a significance of $\gtrsim2\sigma$, while when $A_L$ varies the significance drops to $\gtrsim1\sigma$. This suggest that a significant part of the evidence for higher-$z$ phantom-like behavior in these parametrizations is a consequence of the excess smoothing (relative to what is expected in the best-fit cosmological model) observed in some of the Planck CMB anisotropy multipoles.

\section{Conclusion}
\label{sec:Conclusion}

Adding to the results of Refs.\ \citen{Chan-GyungPark:2024mlx, Chan-GyungPark:2024brx}, here we analyze three new $w(z)$CDM ($+A_L$) dynamical dark energy parametrizations. From where the $\Lambda$CDM model point lies relative to the P18+lensing+non-CMB data cosmological parameter constraint contours of the four $w(z)$CDM ($+A_L$) parametrizations, we find that dark energy dynamics is favored over a cosmological constant by $\gtrsim 2\sigma$ when $A_L = 1$, but only by $\gtrsim 1\sigma$ when $A_L$ is allowed to vary (and is simultaneously determined from these data to be $>1$ at $\sim 2\sigma$ significance). The non-CMB data compilation we use is the largest such compilation of independent, mutually-consistent non-CMB data \cite{deCruzPerez:2024abc}. It is the dominant part of the P18+lensing+non-CMB compilation at low $z$ when these data favor quintessence-like dark energy dynamics. At high $z$, when P18+lensing data are dominant, these data favor phantom-like dark energy dynamics, at significance ranging from $1.5\sigma$ to $2.9\sigma$ when $A_L = 1$ and at a reduced significance ranging from $1.1\sigma$ to $1.8\sigma$ when $A_L$ is allowed to vary and be simultaneously determined from these data. 

From $\Delta$DIC values relative to the $\Lambda$CDM model, we find the $w(z)$CDM parametrizations are {\it positively} favored over the $\Lambda$CDM model, for both the $A_L =1$ (with $\Delta$DIC values ranging from $-2.45$ to $-3.98$) and the varying $A_L$ (with $\Delta$DIC values ranging from $-4.37$ to $-5.60$) cases, with the varying $A_L$ case {\it weakly} to {\it positively} favored over the $A_L = 1$ case (with relative $\Delta$DIC values ranging from $-1.62$ to $-2.15$), because allowing $A_L$ to vary results in better reconciliation of the P18 or P18+lensing data constraints and the non-CMB data constraints.

This evidence for dark energy dynamics in the $w(z)$CDM parametrizations does not depend on the use of DESI data, which we have not used here, also see Refs.\ \citen{Chan-GyungPark:2024mlx, Chan-GyungPark:2024brx}. Nor does it depend on the use of Pantheon+ SNIa (or any SNIa data), see Ref.\ \refcite{Chan-GyungPark:2024mlx} for a detailed study of this in the $w_0w_a$CDM parametrization. The shifts toward quintessence-like behavior of the measured $w(z)$'s, in the $w(z)$CDM ($+A_L$) parametrizations, when $A_L$ is allowed to vary compared to the corresponding $A_L = 1$ case, suggest that this evidence for dark energy dynamics at least partially depends on the observed excess smoothing of some of the Planck CMB anisotropy multipoles, also see Ref.\ \refcite{Chan-GyungPark:2024brx}. In this context, it is interesting that from the new PR4 Planck data release \cite{Tristram:2023haj}, in the flat $\Lambda$CDM$+A_L$ model, where $A_L$ is allowed to vary, PR4 data including lensing data give $A_L = 1.037 \pm 0.37$, favoring $A_L > 1$ at $1\sigma$, which is smaller than the $1.78\sigma$ significance for $A_L > 1$ that follows from the P18+lensing data result $A_L = 1.073 \pm 0.041$.      

While our results are interesting, they are not based on a physically consistent dynamical dark energy model, e.g., $\phi$CDM \cite{Peebles:1987ek, Ratra:1987rm}; they are based on $w(z)$CDM parametrizations that reduce to different XCDM parametrizations at low and high $z$. Our results are also not that statistically significant. However, they highlight a number of interesting issues that deserve additional scrutiny.

\section*{Acknowledgments}
C.-G.P.\ was supported by a National Research Foundation of Korea (NRF) grant funded by the Korea government (MSIT) No.\ RS-2023-00246367. 

\bibliographystyle{ws-ijmpd}
\bibliography{w0w2CDM_IJMPD}

\end{document}